\documentclass[journal]{IEEEtran}

\usepackage{cite}
\usepackage{amsmath,amssymb,amsfonts}
\usepackage{algorithmic}
\usepackage{graphicx}
\usepackage{textcomp}
\usepackage{xcolor,color}
\usepackage{hyperref}
\graphicspath{{figures/}}

\newtheorem{theorem}{Theorem}

\newtheorem{lemma}{Lemma}
\newtheorem{assu}{Assumption}
\newtheorem{pb}{Problem}
\newtheorem{remark}{Remark}
\allowdisplaybreaks
\usepackage{algorithmic}
\def\BibTeX{{\rm B\kern-.05em{\sc i\kern-.025em b}\kern-.08em
    T\kern-.1667em\lower.7ex\hbox{E}\kern-.125emX}}
\begin{document}

\title{Age Optimal Sampling Under Unknown Delay Statistics
\\
\author{Haoyue~Tang,~\IEEEmembership{Student Member,~IEEE},~Yuchao Chen,~Jintao Wang,~\IEEEmembership{Senior Member,~IEEE},~Pengkun Yang, and~Leandros~Tassiulas,~\IEEEmembership{Fellow,~IEEE}
	\thanks{This paper was presented in part IEEE Infocom2022 \cite{Tang2205:Sending}. The work of Y. Chen and J. Wang was supported in part by Tsinghua University-China Mobile Research Institute Joint Innovation Center. The work of H. Tang was supported by the NSF CNS-2112562 AI Institute for Edge Computing Leveraging Next
Generation Networks (Athena). The work of L. Tassiulas was supported by NSF CNS-2112562 and the ONR N00014-19-1-2566. The work of P. Yang was supported by NSFC-12101353 and Tsinghua University Initiative Scientific Research Program. }	
	\thanks{H. Tang was with the Department of Electronic Engineering, Tsinghua University. She is now with the Institute of Network Science, Yale University. (email: haoyue.tang@yale.edu)
}
	\thanks{Y. Chen and J. Wang are with the Department of Electronic Engineering, Tsinghua University, Beijing 100084, China and Beijing National Research Center for Information Science and Technology (BNRist). J. Wang is also with Research Institute of Tsinghua University in Shenzhen, Shenzhen, 518057.  (email: \{cyc20@mails, wangjintao@mail\}.tsinghua.edu.cn)}
	\thanks{P. Yang is with the Center for Statistical Science, Tsinghua University. (email: yangpengkun@tsinghua.edu.cn)}
	\thanks{L. Tassiulas is with the Department of Electrical Engineering and Institute for Network Science, Yale University. (email: leandros.tassiulas@yale.edu)}
	\thanks{(\emph{Corresponding author: Jintao Wang. })}
}
}

\maketitle

\begin{abstract}
This paper revisits the problem of sampling and transmitting status updates through a channel with random delay under a sampling frequency constraint \cite{sun_17_tit}. We use the Age of Information (AoI) to characterize the status information freshness at the receiver. The goal is to design a sampling policy that can minimize the average AoI when the statistics of delay is unknown. We reformulate the problem as the optimization of a renewal-reward process, and propose an online sampling strategy based on the Robbins-Monro algorithm. We prove that the proposed algorithm satisfies the sampling frequency constraint. Moreover, when the transmission delay is bounded and its distribution is absolutely continuous, the average AoI obtained by the proposed algorithm converges to the minimum AoI when the number of samples $K$ goes to infinity with probability 1. We show that the optimality gap decays with rate  $\mathcal{O}\left(\ln K/K\right)$, and the proposed algorithm is minimax rate optimal. Simulation results validate the performance of our proposed algorithm. 
\end{abstract}


\section{Introduction}
With the proliferation of autonomous vehicles and intelligent manufacturing, status updates are becoming a larger part of communications \cite{aoi_intro}. Status updates are crucial to the efficient control and monitoring in such applications, and therefore should be delivered to the destination as timely as possible. To measure the timeliness of status update information at the receiver, the Age of Information (AoI), or simply Age is proposed \cite{roy_12_aoi}. Since then, the design of Age optimal transmission and sampling strategies under communication constraints has received wide attention. 

When the transmission statistics (e.g., delay distribution, packet-loss probabilities) are known in advance, designing AoI minimum transmission strategies can be formulated into a  Markov decision process (MDP) \cite{wang_21_twc,discrete_preempt,bo_sampling,tang_20_jsac,ceran_wcnc,aba_drl_aoi,arafa_online,bedewy_21_tit,sennur_gg1,najm,roy_15_isit,sun_17_tit}. When the generation of status updates are controlled by external sources, AoI minimum cross-layer scheduling and transmission have been studied in \cite{wang_21_twc,discrete_preempt,bo_sampling}; When the generation process can be controlled at will, the joint sampling and transmission of status update packets have been studied in \cite{ceran_wcnc,aba_drl_aoi,arafa_online,tang_20_jsac}. In continuous time scenarios, by modeling the external status update generation as a random process, the expected AoI performance under different service disciplines are analyzed in \cite{bedewy_21_tit,sennur_gg1}. 

Designing AoI minimum sampling strategies in an unknown environment can be formulated as a sequential decision making problem, where online and reinforcement learning algorithms can be employed \cite{aoibandit,atay2020aging, banerjee_adversarial_aoi,tripathi2021online,li2021efficient}. When the generation of status update packets is controlled by external sources, AoI minimum adaptive packet scheduling and link selection algorithms have been proposed \cite{aoibandit,atay2020aging,banerjee_adversarial_aoi}. Tripathi \emph{et al. } model the timeliness of status updates to be a time-varying function of the AoI \cite{tripathi2021online}, and a robust online learning algorithm is proposed. When the status update packets can be generated at will, \cite{li2021efficient} models the data freshness requirement as a minimum AoI constraint, and proposes scheduling algorithms that can achieve a sub-linear utility regret while satisfying the AoI constraint. However, the ultimate goal in \cite{li2021efficient} is to optimize the total utility over the entire network, rather than the AoI performance. Designing Age optimal sampling and transmission strategies have been studied in \cite{ceran_19_infocomwks,ceran_21_jsac,kam_rl,aba_drl_aoi,aylin_rl}, where various deep reinforcement learning algorithms (e.g., SARSA, Actor-Critic, Q-Learning) have been employed. However, the convergence rate of those algorithms are not well understood. Although the online sampling strategies proposed in \cite{chichun,tsai2022distribution} is shown to converge to the optimum strategy almost surely, the optimality of the algorithm is not known. 

In general, although there is a growing number of literature on Age optimal transmission in unknown environment, how to design effective generate-at-will sampling strategies with theoretical guarantees is not well understood. To answer this question, we revisited the point-to-point status update system (Fig.~\ref{fig:model}) in \cite{sun_17_tit,arafa_model}, where a sensor samples and transmits update packets to the destination through a channel with a random delay. The goal is to design an online sampling strategy that minimizes the average AoI at the destination when the delay statistics is unknown. The contributions of the paper are as follows:

\begin{itemize}
	\item Our work is the first to design a Robbins-Monro based online policy to minimize the average AoI when the delay statistics is unknown. Moreover, by using the Lyapunov-Drift-Plus-Penalty approach, our algorithm can satisfy the sampling frequency constraint concurrently (Theorem 5). 
	
	\item When there is no sampling constraint, we show that the time-averaged AoI of the proposed algorithm converges to the limit point of an ordinary differential equation (ODE) almost surely. By showing that the limit point of the ODE is unique and stationary, we prove that the time-averaged AoI obtained by the proposed algorithm converges to the minimum AoI with probability 1 (Theorem 2). The optimality gap of the proposed online learning algorithm decays with rate $\mathcal{O}(\ln K/K)$, where $K$ is the total number of samples (Theorem 3). 
	
	\item By using the Le Cam's two point from non-paramatric statistics, we show that under the worst case delay distribution, the gap between the average AoI of any online learning algorithm and the minimum AoI with known delay statistics decays with rate larger than $\Omega(\ln K/K)$, where $K$ is the total number of samples (Theorem 4). Both the mathematics tool and the converse result is novel in the field of stochastic approximation, and show that the convergence rate of the proposed algorithm (Theorem 3) is minimax order optimum. 
	
\end{itemize}

Independent of this work, \cite{tsai2022distribution} proposes a similar Robbins-Monro algorithm to minimize the average AoI penalty for a two-way delay communication system. It is worth noting that, by using the sampling frequency debt as a dual optimizer, our modified Robbins-Monro algorithm satisfies the sampling frequency constraint at the transmitter side. Our algorithm can be extended to the problem of minimizing the average AoI penalty with a sampling frequency constraint, because computing the optimal updating threshold is equivalent to solving an equation. Moreover, the proof techniques for almost sure convergence are different, with ours using the ODE method. We further establish the minimax lower bound of the average AoI gap of any online algorithm.
\section{Problem Formulation}

\subsection{System Model}
\begin{figure}[h]
	\centering
	\includegraphics[width=.48\textwidth]{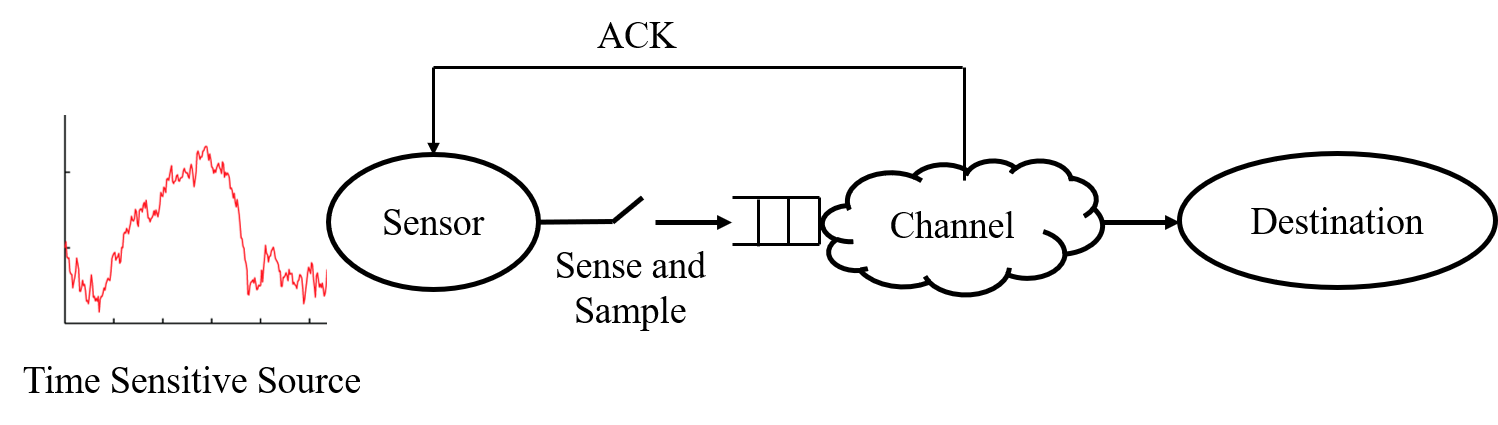}
	\caption{A point-to-point status update system. }
	\label{fig:model}
\end{figure}

Similar to \cite{sun_17_tit,arafa_model}, we consider a status update system depicted in Fig.~\ref{fig:model}, where a sensor observes a time sensitive process, samples status updates and sends them to the destination through a channel. The channel transmits update packets based on a First-Come-First-Serve (FCFS) principle, and each update packet experiences a random transmission delay. Due to the transmission delay, update packets may have to wait in the queue before the last transmission finishes. Once the packet is received by the destination, an acknowledgement (ACK) will be received by the sensor immediately.  

Similar to \cite{roy_15_isit}, suppose the sensor can sample update packets at any time $t\in\mathbb{R}^+$ at his own will. The sampling time-stamp and channel transmission delay of the $k$-th sampled packet are denoted by $S_k$ and $D_k$, respectively. We assume each transmission delay $D_k, k\in\{1, 2, \cdots\}$ is identically and independently distributed (i.i.d.) following the probability measure $\mathbb{P}_D$.
\begin{assu}\label{assu:delay}
	The probability measure $\mathbb{P}_D$ is absolutely continuous on $[0, \infty)$. Its expectation and second order moment is bounded, i.e., 
	\begin{subequations}
		\begin{align}
  &0<\overline{D}_{\mathsf{lb}}\leq\overline{D}\triangleq\mathbb{E}_{\mathbb{P}_D}[D]\leq\overline{D}_{\mathsf{ub}}<\infty,\\ &0<M_{\mathsf{lb}}\leq\mathbb{E}_{\mathbb{P}_D}[D^2]\leq M_{\mathsf{ub}}<\infty. 
		\end{align}
	\end{subequations}
\end{assu}

Let $R_k$ be the reception time-stamp of the $k$-th update packet. Notice that the service of the $k$-th packets starts at $\max\{R_{k-1}, S_k\}$, therefore, $R_k$ can be computed recursively through equation $R_k=\max\{R_{k-1}, S_k\}+D_k$. If the transmission of the $(k-1)$-th update packet has not finished before the $k$-th update packet has been sampled, i.e., $R_{k-1}>S_k$, the $k$-th packet has to wait in the queue and then becomes stale. Therefore, to keep information at the destination fresh, it is better to \emph{wait} for the ACK of the $(k-1)$-th update packet before sampling the $k$-th packet, i.e., $S_k\geq R_{k-1}$. By using such a \emph{waiting} policy, the reception time-stamp of the $k$-th update packet can be simplified to $R_k=S_k+D_k$. We denote $W_k:=S_{k+1}-R_{k}$ to be the \emph{waiting} time after receiving the $k$-th sample. 
\subsection{Age of Information }

AoI measures the time elapsed since the freshest information stored at the destination is generated  \cite{roy_12_aoi}.
Let $i(t):=\arg\max\{k\in\mathbb{N}^+|R_k\leq t\}$ be the index of the latest sample received by the destination before time $t$. 
The AoI at time $t$, denoted by $A(t)$ is:
\begin{equation}
	A(t):=t-S_{i(t)}. 
\end{equation}

\begin{figure}[h]
	\centering
	\includegraphics[width=.48\textwidth]{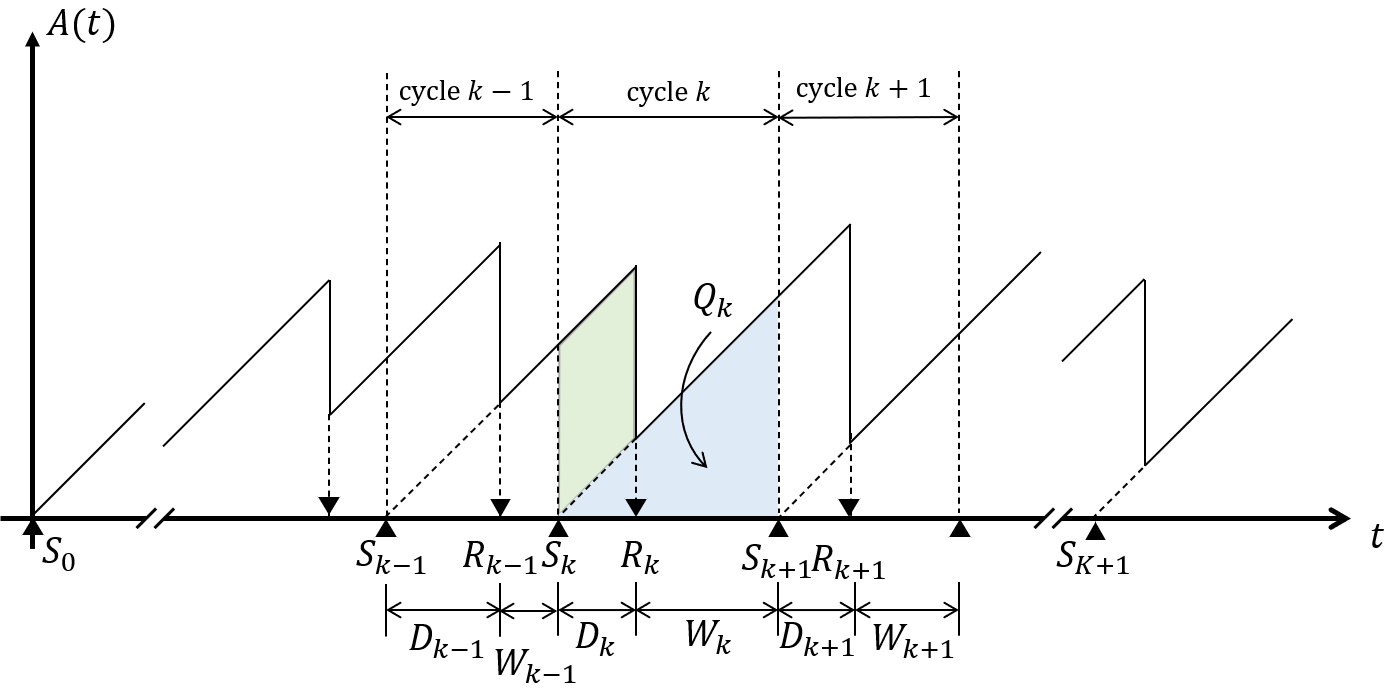}
	\caption{Illustration of AoI evolution.}
	\label{fig:aoievolve}
\end{figure}

A sample path of AoI evolution is depicted in Fig.~\ref{fig:aoievolve}. 

\subsection{Optimization Problem Formulation}
We aim at minimizing the average AoI by designing a sampling strategy $\pi\triangleq\{W_1, W_2, \cdots\}$. Specifically, we only focus on the class of ``\emph{causal}'' policies $\Pi$, where the waiting time $W_k$ is selected based on the past delay and sampling time-stamps denoted by $\mathcal{H}_{k-1}:=\{(S_i, D_i)\}_{i=1}^{k-1}$. No future information $\{D_i\}_{i> k}$ can be used for decision making. To facilitate further analysis, assume that each waiting time is upper bounded by $W_\mathsf{ub}$, and denote $\Pi$ as the class of causal policies whose waiting time $W_k\in[0, W_\mathsf{ub}]$.\footnote{The assumption is reasonable and will not hurt the optimality in policy design when the upper bound $W_{\mathsf{ub}}$ is selected to be large. This is because waiting for an infinitely long time is not beneficial to AoI minimization. } Let $K$ be the total number of sampling times. The expected time average AoI using policy $\pi$ is defined by\footnote{Another definition of the time average AoI can be the limits of expected total AoI over an observation window $[0, T)$ divide the length of the window $T$, i.e., $\limsup_{T\rightarrow\infty}\frac{1}{T}\mathbb{E}\left[\int_0^TA(t)\text{d}t\right]$. The two definitions are both reasonable. Specifically, when $\pi$ is a stationary randomized policy such that the Markov chain $\{(D_k, W_k)\}$ has only one ergodic class, the two definitions are equal \cite{ross2013applied}. }:
\begin{equation}\overline{A}_{\pi}\triangleq\limsup\limits_{K\rightarrow\infty}\frac{\mathbb{E}\left[\int_{t=0}^{S_{K+1}}A(t)\text{d}t\right]}{\mathbb{E}[S_{K+1}]}, \label{eq:avgAoIdef}
\end{equation}
where the AoI $A(t)$ is determined by both the transmission delay $\{D_k\}$ and sampling strategy $\pi$. 

To facilitate further computation and analysis, we define ``\emph{cycle}'' $k$ to be the time interval between the $k$-th and the $(k+1)$-th sampling time-stamps. Since the transmission delay $D_k$ in each cycle $k$ is i.i.d., we have $\mathbb{E}[S_{K+1}]=\mathbb{E}\left[\sum_{k=1}^{K}(D_k+W_k)\right]$. Similarly, let $X_k:=\int_{t=S_k}^{S_{k+1}}A(t)\text{d}t$ be the cumulative AoI in cycle $k$, which is the sum of the area of a parallelogram and a triangle, i.e., 
\[X_k=(D_{k-1}+W_{k-1})D_k+\frac{1}{2}(D_k+W_k)^2.\]

Then the cumulative AoI over interval $[0, S_{K+1})$ can be rewritten as a sum of $X_k$, i.e., \begin{align}
&\mathbb{E}\left[\int_{t=0}^{S_{K+1}}A(t)\text{d}t\right]=\mathbb{E}\left[\sum_{k=1}^KX_k\right]\nonumber\\
=&\mathbb{E}\left[\sum_{k=1}^Kq(D_{k-1}, W_{k-1}, D_k, W_k)\right],\label{eq:aoidecomp}
\end{align} 
where function $q$ is defined as follows:
\[q(d', w', d, w):=(d'+w')d+\frac{1}{2}(d+w)^2. \]

Designing the optimum strategy $\pi$ that minimizes the expected average AoI can be formulated as the following optimization problem:
\begin{pb}{\label{pb:rr-reformulate}}
	\begin{subequations}
		\begin{align}
		\mathsf{AoI}_{\mathsf{opt}}\triangleq&\inf_{\pi \in \Pi}\limsup_{K\rightarrow\infty}\frac{\mathbb{E}\left[\sum_{k=1}^Kq(D_{k-1}, W_{k-1}, D_k, W_k)\right]}{\mathbb{E}\left[\sum_{k=1}^K(D_k+W_k)\right]},\label{eq:primalobj}\\
		&\text{s.t. }\liminf\limits_{K\rightarrow\infty}\frac{1}{K}\mathbb{E}\left[\sum_{k=1}^K(D_k+W_k)\right]\geq\frac{1}{f_{\mathsf{max}}}, \label{eq:samplecons}
		\end{align}
	\end{subequations}
\end{pb}
where $f_{\mathsf{max}}$ is the maximum time average sampling frequency the status update system can afford due to various resource constraints (i.e., energy or system operation frequency). 

Let $\pi^\star$ be the optimum policy that achieves $\mathsf{AoI}_{\mathsf{opt}}$. According to \cite{sun_17_tit}, policy $\pi^\star$ has a threshold structure. When the delay distribution $\mathbb{P}_D$ is known, Sun \emph{et al.} \cite{sun_17_tit} proposed to compute the optimum threshold through a bi-section search. In this paper, we assume only the lower and upper bounds of the average delay and second order moment $\overline{D}_{\mathsf{lb}}, \overline{D}_{\mathsf{ub}}, M_{\mathsf{lb}}$ and $M_\mathsf{ub}$ can be used at the transmitter\footnote{This assumption is reasonable since $\overline{D}_{\mathsf{lb}}$ and $M_{\mathsf{lb}}$ can be computed using the header time, and $\overline{D}_{\mathsf{ub}}$, $M_{\mathsf{ub}}$ can be computed using the maximum Round Trip Time (RTT). }. The closed form expression of distribution $\mathbb{P}_D$ is not accessible to the transmitter and hence cannot be used for decision making.

\section{Problem Resolution}
In this section, we will first reformulate Problem~\ref{pb:rr-reformulate} into a renewal-reward process. 
In Section~\ref{sec:proposealg}, we then propose an adaptive sampling strategy that can learn the optimum policy $\pi^\star$ when the number of samples goes to infinity. The theoretical performance of the algorithm is analyzed in Section~\ref{sec:theoreticana}. 
\subsection{A Renewal-Reward Process Reformulation}
A policy $\pi\in\Pi$ is \emph{stationary deterministic} if the waiting time $W_k$ is a stationary mapping from the transmission delay $D_k$, i.e.,  $W_k=w(D_k)$ and function $w:[0, \infty)\mapsto[0, W_{\mathsf{ub}}]$ is a deterministic function that specifies the waiting time. Let $\Pi_{\mathsf{SD}}\subseteq\Pi$ be the set of stationary deterministic policy such that:
\[\Pi_{\mathsf{SD}}\triangleq\{\pi\in\Pi:W_k=w(D_k), \forall k\}. \]

When $\mathbb{P}_D$ is known, we then have the following theorem according to \cite{sun_17_tit}:

\begin{theorem}{\cite[Theorem 2 Restated]{sun_17_tit} }\label{thm:sd}
	There is a stationary deterministic policy $\pi^\star\in\Pi_{\mathsf{SD}}$ that is optimal to Problem~\ref{pb:rr-reformulate}. 
\end{theorem} 

With slight abuse of notations, we denote $\pi(d)$ to be the waiting time selection function of a stationary deterministic policy by observing transmission delay $d$. With Theorem~\ref{thm:sd}, denote $L_2$ to be the Lebesgue space. Searching for the optimum stationary deterministic policy $\pi^\star$ that achieves $\mathsf{AoI}_{\mathsf{opt}}$ in Problem~\ref{pb:rr-reformulate} can be reformulated into Problem~\ref{pb:primal-rr} as follows:
\begin{pb}[Renewal-Reward Process Optimization Reformulation]\label{pb:primal-rr}
	\begin{subequations}
	\begin{align}
	\mathsf{AoI}_{\mathsf{opt}}=&\inf_{\pi\in L_2}\left(\frac{\mathbb{E}[\frac{1}{2}\left(D+\pi(D)\right)^2]}{\mathbb{E}[D+\pi(D)]}+\overline{D}\right),\label{eq:rrobj}\\
	&\text{s.t. }\mathbb{E}\left[D+\pi(D)\right]\geq\frac{1}{f_{\mathsf{max}}}. 
	\end{align}
	\end{subequations}
\end{pb}

The detailed derivation is the same as \cite{sun_17_tit} and is hence omitted. 
Problem~\ref{pb:primal-rr} can be viewed as the optimization of a renewal-reward process in the sense that:
\begin{itemize}
	\item  The delay $D_k$ observed in each cycle $k$ is i.i.d. following distribution $\mathbb{P}_D$. 
	
	\item  Let $L_k:=D_k+\pi(D_k)$ be the length of the $k$-th cycle. Since $D_k$ is i.i.d. and $\pi(\cdot)$ is a deterministic function, $L_k$ is an i.i.d. random variable. 
	
	\item  Denote $Q_k:=\frac{1}{2}\left(D_k+\pi(D_k)\right)^2$, which can be viewed as the reward received in cycle $k$. Due to the i.i.d. assumption of $D_k$, the reward $Q_k$ is also an i.i.d. random variable. 
\end{itemize}

As a result, the length and reward $(L_k, Q_k)$ in frame $k$ is independent of $(L_{k'}, Q_{k'})$ in other frames $k'\neq k$. Moreover, the expectation $\mathbb{E}[L_k]\leq\mathbb{E}[D+W_{\mathsf{ub}}]<\infty$ and $\mathbb{E}[Q_k]\leq\mathbb{E}[\frac{1}{2}(D+W_{\mathsf{ub}})^2]<\infty$ are both bounded. Problem~\ref{pb:primal-rr} cast into the renewal-reward process optimization framework.

\subsection{Proposed Online Algorithm}\label{sec:proposealg}
We will first review the computation of $\pi^\star$ when the delay statistics $\mathbb{P}_D$ is known, and then propose an online algorithm that learns policy $\pi^\star$ adaptively. For simplicity, let $\Pi_{\mathsf{cons}}$ be the set of stationary deterministic policies whose sampling frequency is below $f_{\mathsf{max}}$, i.e., 
\[\Pi_{\mathsf{cons}}\triangleq\{\pi\in\Pi_{\mathsf{SD}}| \mathbb{E}\left[D+\pi(D)\right]\geq\frac{1}{f_\mathsf{max}}\}.\]

\subsubsection{Design $\pi^\star$ with known $\mathbb{P}_D$}
Recall that $\overline{A}_{\pi^\star}$ is the minimum time average AoI any policy $\pi\in\Pi_{\mathsf{cons}}$ can achieve, i.e., 
\begin{equation}\overline{A}_\pi=\frac{\mathbb{E}\left[\frac{1}{2}(D+\pi(D))^2\right]}{\mathbb{E}[D+\pi(D)]}+\overline{D}\geq\overline{A}_{\pi^\star}. \label{eq:ageq}
\end{equation}
Deducting $\overline{D}$ on both sides of inequality \eqref{eq:ageq}, we have:
\begin{equation}
    \frac{\mathbb{E}\left[\frac{1}{2}(D+\pi(D))^2\right]}{\mathbb{E}[D+\pi(D)]}\geq\overline{A}_{\pi^\star}-\overline{D}. \forall \pi\in\Pi_{\mathsf{cons}}.\label{eq:renewalrewartopt}
\end{equation}

For simplicity, denote $\gamma^\star=\overline{A}_{\pi^\star}-\overline{D}$ and then then multiplying $\mathbb{E}[D+\pi(D)]$ on both sides of inequality \eqref{eq:renewalrewartopt}, we then have the following inequality:
\begin{equation} \frac{1}{2}\mathbb{E}[(D+\pi(D))^2]-\gamma^\star\mathbb{E}[D+\pi(D)]\geq 0, \forall \pi\in\Pi_{\mathsf{cons}}. \label{eq:opt-inequality}\end{equation}

Notice that \eqref{eq:opt-inequality} takes equality if and only if policy $\pi$ is AoI minimum. Therefore, when $\gamma^\star$ is known, $\pi^\star$ can be obtained by solving the following functional optimization problem:
\begin{pb}[Functional Optimization Problem]\label{pb:c-rr-problem}
	\begin{subequations}
\begin{align}
\theta_{\mathsf{opt}}\triangleq&\mathop{\min}\limits_{\pi\in\Pi_{\mathsf{SD}}}\mathbb{E}\left[\frac{1}{2}(D+\pi(D))^2-\gamma^\star(D+\pi(D))\right],\label{eq:consopt}\\
&\hspace{0.9cm}\text{s.t. }\mathbb{E}[D+\pi(D)]\geq\frac{1}{f_{\mathsf{max}}}. \label{eq:freqcons}
\end{align}
\end{subequations}
\end{pb}

Inequality \eqref{eq:opt-inequality} shows $\theta_{\mathsf{opt}}=0$. To find the optimum policy that achieves $\theta_{\mathsf{opt}}$, we place the sampling frequency constraint \eqref{eq:freqcons} into the objective function \eqref{eq:consopt} using a dual optimizer $\underline{\nu\geq0}$, we can formulate the Lagrange function as follows:
\begin{align}
	\mathcal{L}(\gamma, \nu, \pi):=&\mathbb{E}\left[\frac{1}{2}(D+\pi(D))^2-(\gamma+\nu)(D+\pi(D))\right]\nonumber\\
 &+\nu\frac{1}{f_{\mathsf{max}}}. \label{eq:lagrange}
\end{align}

 As is shown in \cite[Theorem 4]{sun_17_tit}, for fixed $\gamma$ and $\nu$, the optimum policy $\pi_{\gamma, \nu}^\star$ that minimizes the Lagrange function \eqref{eq:lagrange} specifies the waiting time through:
\begin{equation}
	\pi_{\gamma, \nu}^\star(d)=(\gamma+\nu-d)^+. \label{eq:piopt}
\end{equation}

Plugging the optimum policy into the Lagrange function \eqref{eq:lagrange}, we have:
\begin{align}
    &\inf_\pi\mathcal{L}(\gamma, \nu, \pi)\nonumber\\
    =&\mathbb{E}\left[\frac{1}{2}\max\{(\gamma+\nu), D\}^2-\gamma\max\{\gamma+\nu, D\}\right]\nonumber\\
    &+\nu\left(\frac{1}{f_\mathsf{max}}-\mathbb{E}\left[\max\{(\gamma+\nu), D\}\right]\right). 
\end{align}

Let $\nu^\star:=\mathop{\arg\sup}\limits_{\nu\geq 0}\inf_{\pi\in\Pi_{\mathsf{SD}}}\mathcal{L}(\gamma^\star, \nu, \pi)$ be the dual optimizer that resolves the Lagrange function when $\gamma=\gamma^\star$. 
Notice that when $\pi^\star=\pi_{\gamma^\star, \nu^\star}^\star$ is used, 
\begin{align}&\theta_{\mathsf{opt}}\nonumber\\
=&\mathbb{E}\left[\frac{1}{2}\max\{(\gamma^\star+\nu^\star), D\}^2-\gamma^\star\max\{(\gamma^\star+\nu^\star), D\}\right]\nonumber\\
=&0. 
\end{align}

We then have the necessary condition on $\gamma^\star$:
\begin{align}
    &\mathbb{E}\left[\frac{1}{2}\max\{(\gamma^\star+\nu^\star), D\}^2-\gamma^\star\max\{(\gamma^\star+\nu^\star), D\}\right]\nonumber\\
    =&0.\label{eq:eqn}
\end{align}

The following lemma characterizes the upper and lower bound of $\gamma^\star$, the proof will be provided in Appendix~\ref{pf:gammaub}:
\begin{lemma}\label{coro:gammaub}
	The optimum ratio $\gamma^\star$ can be upper and lower bounded by:
\[\gamma_{\mathsf{lb}}\leq\gamma^\star\leq\gamma_{\mathsf{ub}},\]
where
\begin{align}
	\gamma_{\mathsf{lb}}&:=\frac{1}{2}\overline{D}_{\mathsf{lb}},\nonumber\\
	\gamma_{\mathsf{ub}}&:=\frac{\frac{1}{2}M_{\mathsf{ub}}+\overline{D}_{\mathsf{ub}}\frac{1}{f_{\mathsf{max}}}+\frac{1}{2}\frac{1}{f_{\mathsf{max}}^2}}{\overline{D}_{\mathsf{lb}}+\frac{1}{f_{\mathsf{max}}}}. \nonumber
\end{align}
\end{lemma}

\subsubsection{An online learning algorithm $\pi_{\mathsf{online}}$ through the Robbins-Monro algorithm}
When the delay statistics $\mathbb{P}_D$ is known,  $(\gamma^\star+\nu^\star)$ can be computed directly using a bi-section method \cite{sun_17_tit}. When $\mathbb{P}_D$ is unknown, such computation is impossible because equation \eqref{eq:eqn} is unknown. As an alternative, we approximate $\gamma^\star$ and $\nu^\star$ respectively. To meet the frequency constraint, we use sequence $\{U_k\}$ to track the sampling frequency constraint violation up to time $S_k$. Notice that the use of dual optimizer $\nu$ is to guarantee the sampling frequency constraint is satisfied, we use $\nu_k=\frac{1}{V}U_k$ as the dual optimizer in cycle $k$, where $V>0$ is fixed as a constant. Then to find the root $\gamma^\star$ of equation \eqref{eq:eqn} assuming that $\nu^\star=\nu_k$ is the dual optimizer, we use a sequence $\{\gamma_k\}$ to approximate $\gamma^\star$ in cycle $k$ using the Robbins-Monro algorithm \cite{robbins_monro}. We start by initializing $\gamma_{1}\in\text{Uni}\left([\gamma_{\text{lb}}, \gamma_{\text{ub}}]\right)$. The algorithm operates in cycle $k$ as follows:
\begin{subequations}
\begin{itemize}	

\item After the transmission delay $D_k$ of the $k$-th update packet is observed, we choose a waiting time $W_k$ based on the current estimation $\gamma_k$ and violation $U_k$:
\begin{equation}
    W_{ k}=\left(\gamma_{k}+\frac{1}{V}U_k-D_{ k}\right)^+,\label{eq:waitingeq}
\end{equation}
where $V>0$ is fixed as a constant. We then wait for $W_k$ to take the next sample and then compute the cycle length $L_k=D_k+W_k$ as well as reward $Q_k=\frac{1}{2}(D_k+W_k)^2$. 

\item  We then update $\gamma_{k}$ via the Robbins-Monro algorithm \cite{robbins_monro} as follows:
\begin{align}
&\gamma_{k+1}=\left[\gamma_{k}+\eta_k\left(Q_{k}-\gamma_{k}L_{ k}\right)\right]_{\gamma_{\text{lb}}}^{\gamma_{\text{ub}}},\label{eq:robbins-monro-gamma}
\end{align}where $[\gamma]_{a}^b=\min\{b, \max\{\gamma, a\}\}$ and $\{\eta_k\}$ is a set of diminishing step sizes that is selected to be:
\begin{equation}
	\eta_k=\begin{cases}
	\frac{1}{2\overline{D}_{\text{lb}}}, &k=1;\\
	\frac{1}{(k+2)\overline{D}_{\text{lb}}}, &k\geq 2. 
	\end{cases}
\end{equation}

\item To guarantee that the sampling frequency constraint is not violated, we update the violation $U_k$ up to the end of cycle $k$ using:
\begin{equation}
U_{k+1}=\left(U_k+\left(\frac{1}{f_{\mathsf{max}}}-L_k\right)\right)^+. \label{eq:debt-evolve}
\end{equation}\end{itemize}
\end{subequations}

\subsection{Theoretic Analysis}\label{sec:theoreticana}
The evolution of the time average AoI optimality gap as a function of time $t$ is hard to analyze in general. As an alternative, define ratio \begin{equation}\tilde{A}_K:=\frac{\mathbb{E}\left[\int_{t=0}^{S_{K+1}}A(t)\text{d}t\right]}{\mathbb{E}[S_{K+1}]}.\label{eq:atildedef}
\end{equation}

This metric is reasonable in the sense that $\tilde{A}_K$ is the ratio between the expected cumulative AoI up to the $K$-th cycle and the running length up to cycle $K$. Let $\pi_K$ be the waiting time specification rule in cycle $K$. According to equation \eqref{eq:waitingeq}, function $\pi_K(d)=(\gamma_K+\frac{1}{V}U_K-d)^+$. We measure the performance of the proposed algorithm via the convergence rate of difference $\tilde{A}_K-\overline{A}_{\pi^\star}$ and the expected average AoI difference between using policy $\pi_K$ and $\pi^\star$, i.e.,  $\overline{A}_{\pi_K}-\overline{A}_{\pi^\star}$. The main results are as follows:
\begin{theorem}\label{thm:converge}
	When there is no transmission constraint, i.e., $f_{\mathsf{max}}=\infty$ and the transmission delay $D<B<\infty$ is upper bounded by $B$, by using the proposed online sampling algorithm $\pi_{\mathsf{online}}$, the threshold $\{\gamma_k\}$ converges to the optimum threshold $\gamma^\star$ with probability 1, i.e., 
	\begin{subequations}
	\begin{equation}
		\lim_{K\rightarrow\infty}\gamma_K\overset{\text{a.s.}}{=}\gamma^\star.\label{eq:thm1-1}
	\end{equation}
	
		As a result, the average AoI of the proposed policy converges to the minimum $\overline{A}_{\pi^\star}$ with probability 1, i.e., 
	\begin{equation}
	    \lim_{K\rightarrow\infty}\frac{\int_{0}^{S_{K+1}}A(t)\text{d}t}{S_{K+1}}\overset{\text{a.s.}}{=}\overline{A}_{\pi^\star},\label{eq:theorem4-as-avgaoi}
	\end{equation}
		\end{subequations}
\end{theorem}
    Proof for Theorem~\ref{thm:converge} is provided in Appendix~\ref{pf:as}.

The next theorem characterizes the convergence rate of the proposed algorithm, whose proof is provided in Appendix~\ref{pf:regret}:
\begin{theorem}\footnote{By selecting proper stepsizes, the results still holds if the upper and lower bound on $\gamma^\star$ is unknown \cite{tangmobihoc}}\label{thm:regret}
Up to frame $K$, the difference $\mathbb{E}[(\gamma_{K}-\gamma^\star)^2]$ can be bounded by:
	\begin{subequations}\begin{equation}
    \mathbb{E}[(\gamma_K-\gamma^\star)^2]\leq\frac{1}{K}\frac{L_{\mathsf{ub}}^4}{\overline{D}_{\mathsf{lb}}^2}.\label{eq:theorem4-gammadiff}
\end{equation}

	The difference between the expected time-averaged AoI by using policy $\pi_K$ and $\pi^\star$ can be upper bounded by:

	\begin{align}
	\overline{A}_{\pi_K}-\overline{A}_{\pi^\star}\leq\frac{L_\mathsf{ub}^4}{\overline{D}\overline{D}_{\mathsf{lb}}^2}\frac{1}{K}=\mathcal{O}\left(\frac{1}{K}\right).\label{eq:theorem4-conclusion2}
	\end{align}
	and the difference $\tilde{A}_K-\overline{A}_{\pi^\star}$ can be upper bounded by:
	
	\begin{align}
    \tilde{A}_K-\overline{A}_{\pi^\star}\leq\frac{L_{\mathsf{ub}}^4}{\overline{D}\overline{D}_{\mathsf{lb}}^2}\times \frac{1+\ln K}{K}=\mathcal{O}\left(\frac{\ln K}{K}\right),\label{eq:theorem4-conclusion1}
\end{align}
where $L_{\text{ub}}=B+\gamma_{\text{ub}}$.

	\end{subequations}
\end{theorem}

\begin{remark}
When there is no sampling constraint, the proposed online algorithm learns the optimum policy adaptively, since both $\overline{A}_{\pi_K}-\overline{A}_{\pi^\star}$ and $\tilde{A}_{\pi_K}-\overline{A}_{\pi^\star}$ goes to 0 as $K$ goes to infinity. 
\end{remark}

\begin{remark}
As is shown in equation \eqref{eq:theorem4-gammadiff}-\eqref{eq:theorem4-conclusion1}, if the estimated average transmission lower bound $\overline{D}_{\text{lb}}$ is closer to $\overline{D}$ and the upper bound $L_{\mathsf{ub}}$ is closer to $\overline{L}^\star$, the upper bound of both the estimation error $\mathbb{E}[(\gamma_K-\gamma^\star)^2]$ and the average AoI difference $\overline{A}_K-\overline{A}_{\pi^\star}$ are be smaller. This implies a good estimation on the upper and lower bound of $\overline{D}$ help minimize the average AoI. 
\end{remark}

\begin{theorem}\label{thm-converse}
	Let $\pi^\star_{\mathbb{P}}$ denote the AoI minimum sampling policy when the delay distribution is $\mathbb{P}$ and let $\gamma^\star_{\mathbb{P}}$ be the optimum updating threshold. At the end of cycle $k$, let $\hat{\gamma}:\mathbb{R}^k\mapsto\mathbb{R}^+$ be an estimator of ratio $\gamma_{\mathbb{P}}^\star$ using historical transmission delays $\mathcal{H}_k$. The minimax estimation error of $\gamma_{\mathbb{P}}^\star$ satisfies:
	\begin{equation}
	    \min_{\hat{\gamma}}\max_{\mathbb{P}}\mathbb{E}\left[(\hat{\gamma}(\mathcal{H}_k)-\gamma_{\mathbb{P}}^\star)^2\right]\geq\Omega(1/k). \label{eq:gammahat}
	\end{equation}
	For any $\delta$ satisfies $0<\delta<\left(\sqrt[3]{\frac{1}{2}+\sqrt{\frac{5}{4}}}+\sqrt[3]{\frac{1}{2}-\sqrt{\frac{5}{4}}}\right)/2$, let $\mathcal{P}_{w}(\delta)$ be the set of delay distributions that: (i) is absolutely continuous and upper bounded by $B$; (ii) when delay $D\sim\mathbb{P}$, by using the AoI optimum policy $\pi_{\mathbb{P}}^\star$, the probability of waiting to take the next sample is larger than $\delta$, i.e., $p_w(\mathbb{P}):=\mathbb{E}_{D\sim \mathbb{P}}\left[\text{Pr}(D\geq\gamma_{\mathbb{P}}^\star)\right]\geq\delta$. Then the time average AoI using any causal sampling algorithm $\pi$ has the following lower bound:
	\begin{equation}
	    \inf_{\pi\in\Pi}\sup_{\mathbb{P}\in\mathcal{P}_w(\delta)}\left(\frac{\mathbb{E}\left[\int_0^{S_{K+1}}A(t)\text{d}t\right]}{\mathbb{E}[S_{K+1}]}-\overline{A}_{\pi^\star_{\mathbb{P}}}\right)\geq\delta\cdot\Omega\left(\frac{\ln K}{K}\right).\label{eq:avgAoIconverse}
	\end{equation}
\end{theorem}

The proof is provided in Appendix~\ref{pf:thm-converse}. 

\begin{remark}
The order of the convergence rate of the $\mathbb{E}[(\gamma_k-\gamma^\star)^2]$ and $\mathbb{E}[\tilde{A}_k-\overline{A}_{\pi^\star}]$ (Theorem~\ref{thm:regret}) match the converse bounds in Theorem~\ref{thm-converse}. Therefore, the proposed algorithm is minimax order optimal. 
\end{remark}

Next, we analyze the sampling frequency violation behaviour of the proposed online policy. We have the following assumptions:
\begin{assu}\label{assu:strictlyfeasible}
	Problem~\ref{pb:rr-reformulate} can be strictly feasible.  There exists $\epsilon>0$ and a $\pi_\epsilon \in\Pi_{\mathsf{SD}}$ so that, by using policy $\pi_\epsilon$, we have the following inequality, 
	\begin{equation}
		\mathbb{E}\left[D+\pi(D)\right]\geq\frac{1}{f_{\mathsf{max}}}+\epsilon. 
	\end{equation}
\end{assu}

Under Assumption~\ref{assu:strictlyfeasible}, we have the following result:
\begin{theorem}\label{thm:violation}
	The sampling constraint can be satisfied in the sense that:
	\begin{equation}
		\liminf\limits_{K\rightarrow\infty}\mathbb{E}\left[\frac{1}{K}\sum_{k=1}^K(W_k+D_k)\right]\geq\frac{1}{f_{\mathsf{max}}}. 
	\end{equation}
\end{theorem}

The proof is provided in Appendix \ref{pf:violation}.

\section{Simulation Results}
We validate the performance of the proposed algorithms via numerical simulations. We consider two sets of heavy tailed distribution that characterize the heavy traffic characteristics: 
\begin{itemize}
    \item[(a)] $\mathsf{lognormal}(\mu, \sigma)$: log-normal distribution parameterized by $\mu$ and $\sigma$, i.e., the density function of the transmission delay distribution is $
p(x)=\frac{1}{\sigma\sqrt{2\pi}}\exp\left(-\frac{(\ln x-\mu)^2}{2\sigma^2}\right). $ 
\item[(b)] $\mathsf{Weilbur}(a, b)$: Weilbur distribution parameterized by scale parameter $a$ and shape parameter $b$, i.e., the density function $p(x)=\frac{b}{a}\left(\frac{x}{a}\right)^{b-1}\exp\left(-\left(\frac{x}{a}\right)^b\right)$. 
\end{itemize}

\subsection{Updating without a Sampling Frequency Constraint}
\begin{figure}[h]
	\centering
	\includegraphics[width=.45\textwidth]{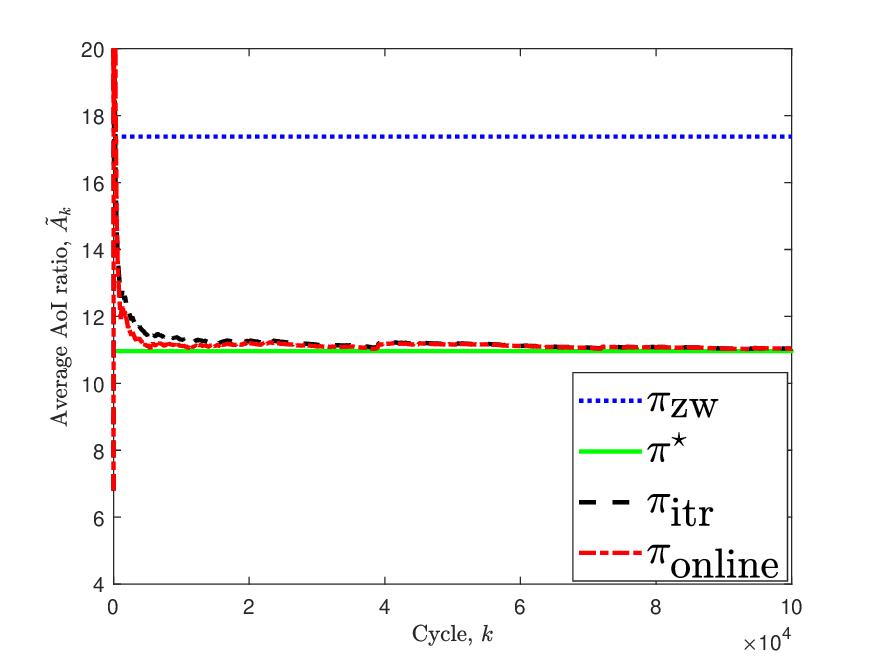}
	\includegraphics[width=.45\textwidth]{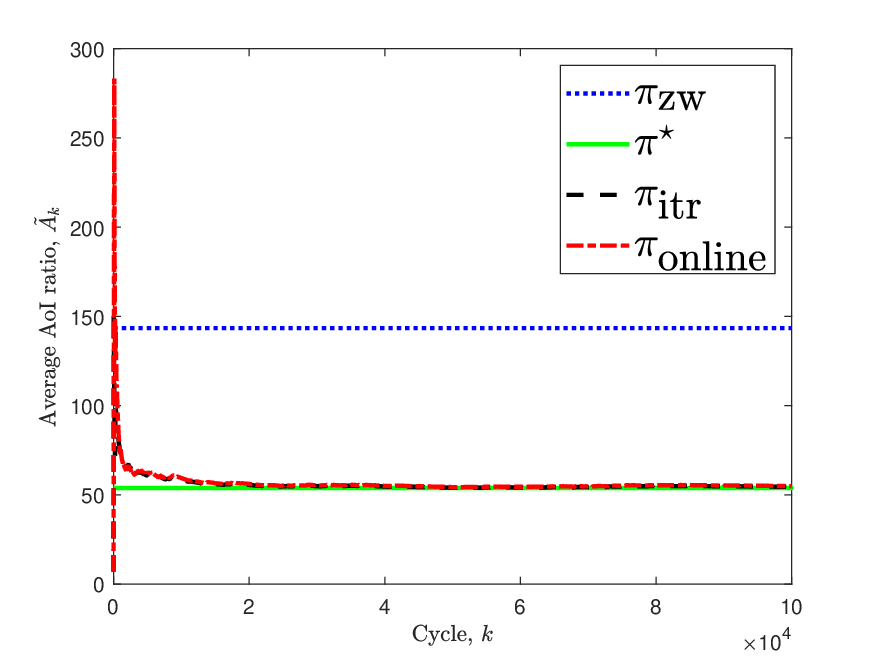}
	\caption{The average AoI ratio evolution as a function of cycle $k$. Left $\mathsf{lognormal}(1, 1.3)$; Right $\mathsf{Weibul}(1, 0.3)$}
	\label{fig:frameasympto}
\end{figure}

\begin{figure}[h]
    \centering
    \includegraphics[width=.45\textwidth]{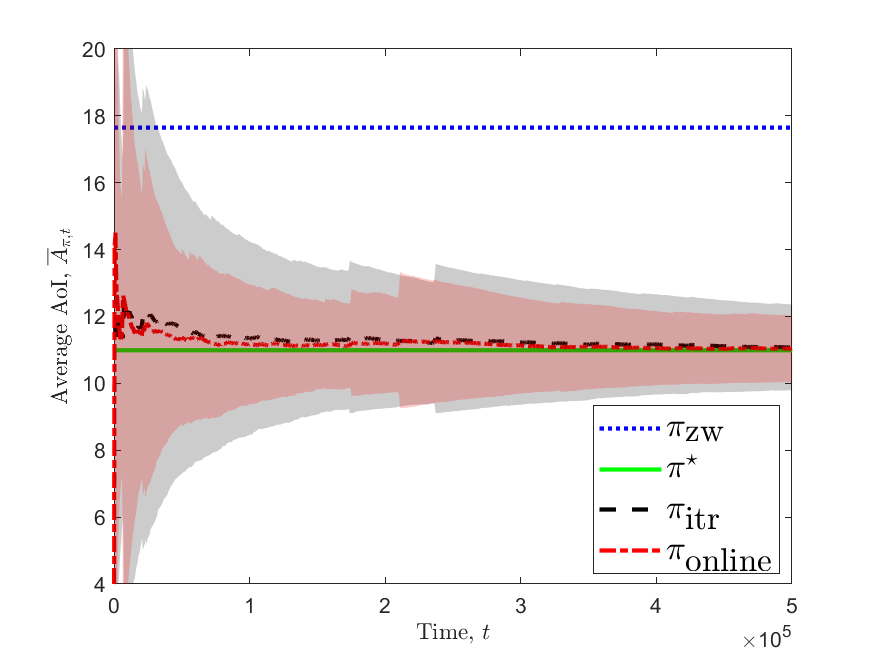}
    \includegraphics[width=.45\textwidth]{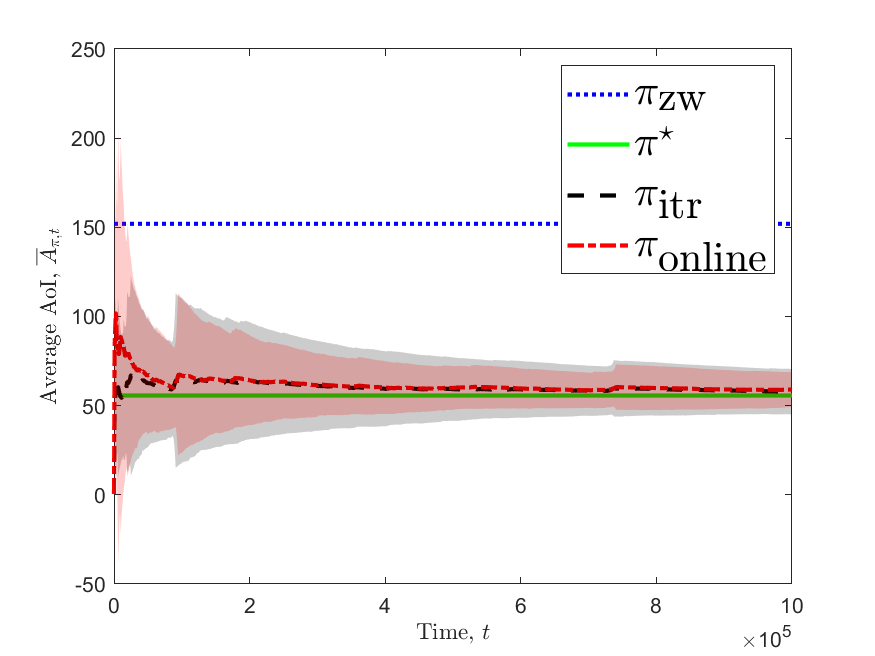}
    \caption{The average time AoI evolution. Left $\mathsf{lognormal}(1, 1.3)$; Right $\mathsf{Weibul}(1, 0.3)$}
    \label{fig:timeasympto}
\end{figure}

We first verify the asymptotic performance of $\pi_{\mathsf{online}}$ when there is no sampling frequency constraint, i.e., $f_{\mathsf{max}}=\infty$. We study and compare the following three strategies: (1) zero-wait policy that specifies $\pi_{\mathsf{zw}}(d)=0, \forall d$; (2) the optimum policy $\pi^\star$ computed by \cite{sun_17_tit}; (3) the iterative threshold computation method $\pi_{\mathsf{itr}}$ proposed by \cite{chichun}. We compute the empirical mean and second-order moment of the first 100 transmission delays, i.e., $\hat{D}=\frac{1}{100}\sum_{k=1}^{100}D_k$, $\hat{M}=\frac{1}{100}\sum_{k=1}^{100}D_k^2$. We then set $D_{\mathsf{lb}}=\hat{D}/10, D_{\mathsf{ub}}=10\hat{D}$ $M_{\mathsf{lb}}=\hat{M}/10, M_{\mathsf{ub}}=10\hat{M}$. Simulations are carried out when the transmission delay follows the log-normal distribution with parameters $\mu=1$ and $\sigma=1.3$. We plotted the AoI ratio up to cycle $k$, i.e.,  $\tilde{A}_k=\frac{\mathbb{E}\left[\int_0^{S_{k+1}}A(t)\text{d}t\right]}{\mathbb{E}[S_{K+1}]}$ in Fig.~\ref{fig:frameasympto}. The mean of the time average AoI $\overline{A}_{\pi, t}=\frac{1}{t}\int_{t'=0}^tA(t')\text{d}t'$ as well as its confidence interval are illustrated in Fig.~\ref{fig:timeasympto}. All the expectations are computed by taking the average of 100 runs. According to Fig.~\ref{fig:frameasympto}, the AoI ratio $\tilde{A}_k$ converges to the optimum AoI obtained by the optimum policy $\pi^\star$, which has been proved theoretically in Theorem~\ref{thm:regret}. Moreover, when the proposed online learning policy $\pi_{\mathsf{online}}$ is used, the optimality gap between $\overline{A}_{\pi, t}$ AoI and the minimum AoI $\overline{A}_{\pi^\star}$ diminishes when time $t$ goes to infinity. Compared with policy $\pi_{\mathsf{itr}}$, the average AoI ratio of our proposed algorithm converges faster to $\overline{A}_{\pi^\star}$ and the variance is smaller. 

\subsection{Updating under a Sampling Frequency Constraint}
\begin{figure*}[h]
	\centering
	\includegraphics[width=.9\textwidth]{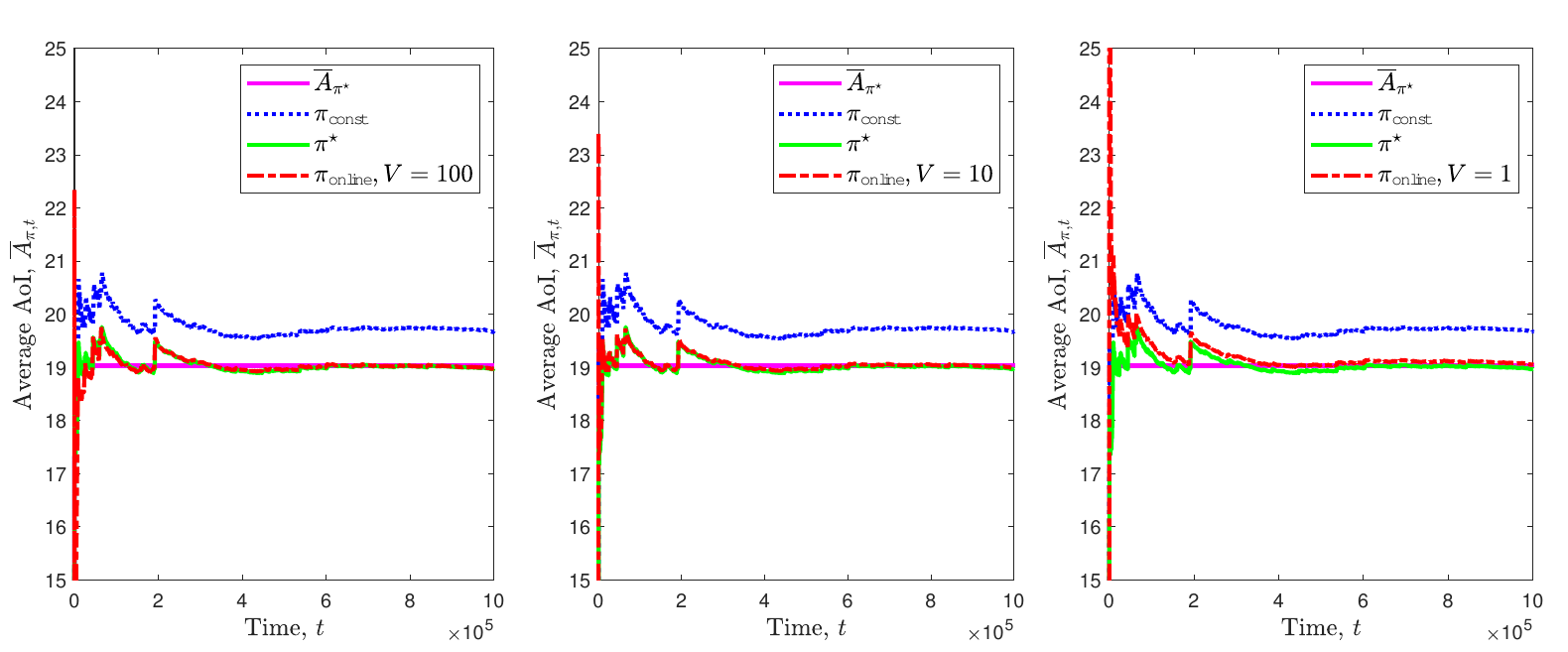}
	\caption{The average AoI ratio evolution of a single sample path under sampling constraint. }
	\label{fig:consAoI}
\end{figure*}

\begin{figure*}[h]
    \centering
    \includegraphics[width=.9\textwidth]{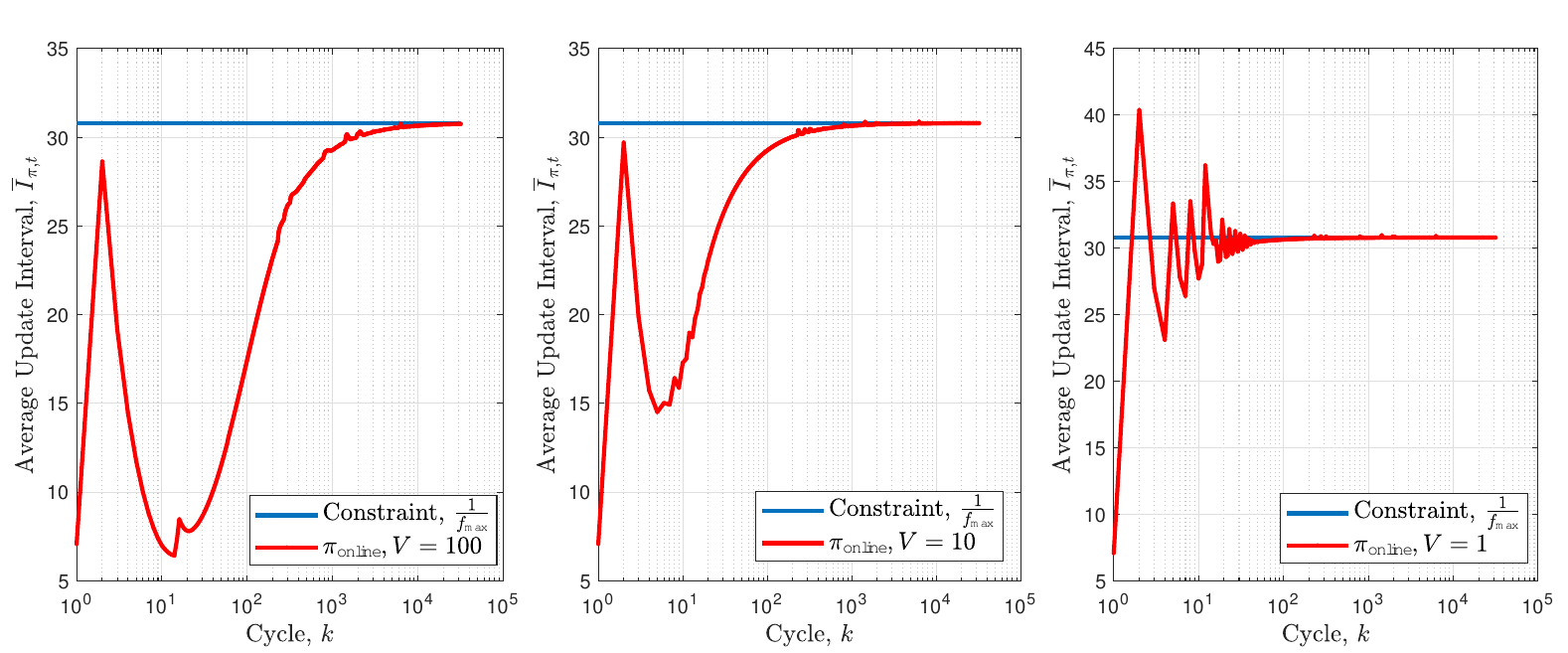}
    \caption{The average sampling interval of a single sample path using different $V$. }
    \label{fig:consfreq}
\end{figure*}

Next we study the performance of the proposed algorithm when the sampling constraint exists. Since the zero-wait sampling policy and the iterative threshold computing policy \cite{chichun} may not satisfy the sampling frequency constraint, we compare the proposed algorithm with (1) a constant wait policy $\pi_{\mathsf{const}}$ that specifies waiting time by $\pi_{\mathsf{const}}(d)=\frac{1}{f_{\mathsf{max}}}-\overline{D}, \forall d$; (2) the optimum policy $\pi^\star$ computed by \cite{sun_17_tit}. Simulations are carried out when the transmission delay follows the log-normal distribution with parameter $\mu=1$, $\sigma=1.5$, and the sampling frequency constraint is selected to be $f_{\mathsf{max}}=\frac{1}{10\overline{D}}$. We plot the average AoI performance of a single sample path in Fig.~\ref{fig:consAoI} and the corresponding average sampling interval $\overline{I}_{\pi, K}\triangleq\frac{S_{K+1}}{K}$ in Fig.~\ref{fig:consfreq}. From Fig.~\ref{fig:consAoI}, it can be observed that the constant wait policy incurs a larger AoI, which is harmful to the data freshness performance. As expected, the average AoI of the proposed online algorithm converges to the average AoI of the optimum policy $\pi^\star$ when time $t$ goes to infinity. Moreover, when time $t$ increases, the average sampling interval converges to $\frac{1}{f_{\mathsf{max}}}$, which means the sampling frequency is not violated. Similar to the queueing length-utility trade-off in network utility maximization \cite{neelydpp}, we found that choosing a smaller $V$ (i.e., $V=1$ in Fig.~\ref{fig:consfreq}) guarantees that the sampling frequency constraint can be satisfied at a earlier stage, while choosing a larger $V$ (i.e., $V=100$ or $V=10$ in Fig.~\ref{fig:consAoI} shows that the average AoI converges to the minimum AoI faster.

\subsection{Addressing Practical Issues in Communication Networks--Timeout}
Preemption, i.e., stop the previous transmission and restart a new on when the transmission delay is larger than a threshold can effectively minimize the average AoI. 
As is revealed by \cite[Lemma 1]{arafa_model}, for pre-emption strategies with threshold $\tau$, i.e., take a new sample and transmit it when the previous delay is larger than $\tau$, the optimum sampling strategy $\pi_{\mathsf{pre}}^{\tau, \star}$ still has a threshold structure. Let $n_k$ be the number of retransmissions before the ACK of the $(k-1)$-th received sample and let $D_k$ be the transmission delay of the $(k-1)$-th received sample, after the ACK of the $(k-1)$-th sample is received, policy $\pi_{\mathsf{pre}}^{\tau, \star}$ selects waiting time $W_k$ as follows:
\begin{equation}
    W_k=(\gamma^\star_{\mathsf{pre}}+\nu^\star_{\mathsf{pre}}-\tilde{D}_k)^+,
\end{equation}
where $\tilde{D}_k:=n_k\tau+D_k$ and the coefficient $\gamma^\star_{\mathsf{pre}}=\overline{A}_{\pi_{\mathsf{pre}}^{\tau, \star}}-\mathbb{E}[D|D\leq\tau]$ is defined similar to $\gamma^\star$, $\nu^\star_{\mathsf{pre}}$ is the dual optimizer for satisfying the sampling frequency constraint. For threshold policies with transmission preemption, the length of frame $k$ now becomes $L_k$ and the reward becomes $Q_k=\frac{1}{2}L_k^2$. Plugging the computation of $L_k$ and $Q_k$ back into algorithm \eqref{eq:waitingeq}-\eqref{eq:debt-evolve} yields the online algorithm with transmission preemption. 

\begin{figure}[h]
    \centering
    \includegraphics[width=.45\textwidth]{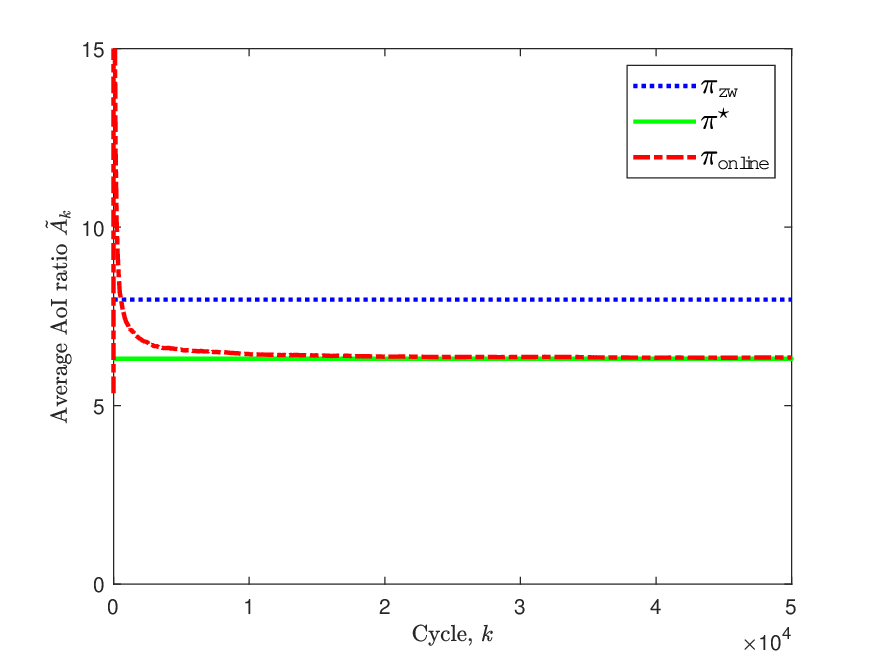}
    \caption{The average AoI performance with time-out. }
    \label{fig:cutoff}
\end{figure}

In Fig.~\ref{fig:cutoff}, we plotted the average AoI of different algorithms when a timeout threshold of $\tau=10$ is used. The transmission delay follows $\mathsf{lognormal}(1, 1.3)$. From Fig.~\ref{fig:cutoff}, the average AoI of our proposed online learning algorithm achieves a smaller AoI compared with the zero-wait policy, and approaches the optimum when the number of samples approaches infinity. 

\section{Conclusions}
In this paper, we considered a sensor sampling and transmitting status updates to the receiver over a channel with random delay. We addressed the problem of minimizing the expected time average AoI under a sampling frequency constraint when the delay distribution is unknown. We reformulated the AoI minimization problem into a renewal-reward process optimization, and we propose an online sampling strategy based on the Robbins-Monro algorithm. We proved that the proposed algorithm can learn the optimum sampling policy almost surely when the number of samples $K$ goes to infinity, and the average sampling frequency constraint can be satisfied. We prove that the convergence rate of the proposed algorithm is minimax optimum under certain conditions. Simulation results validate the adaptive performance of the proposed algorithm. Interesting extensions with piece-wise stationary delay distribution will be our future work. 

\section*{Acknowledgement}
It is a pleasure to thank Prof. Yin Sun for pointing out the ODE approach for the proof of Theorem~\ref{thm:converge} and for insightful suggestions on Theorem~\ref{thm-converse},  Prof. Yuhao Wang for discussions on the proof of Theorem~\ref{thm:esterror} and Dr. Hengjie Yang for a careful proofreading of the draft. 

\appendices

\section{Proof of Lemma~\ref{coro:gammaub}}\label{pf:gammaub}
\begin{IEEEproof}The lower bound of $\gamma^\star$ can be computed as follows:
\begin{align}
	\gamma^\star&=\frac{\mathbb{E}\left[\frac{1}{2}(D+\pi^\star(D))^2\right]}{\mathbb{E}[D+\pi^\star(D)]}\nonumber\\
	&\overset{(a)}{\geq}\frac{1}{2}\frac{\mathbb{E}[D+\pi^\star(D)]^2}{\mathbb{E}[D+\pi^\star(D)]}\nonumber\\
	&=\frac{1}{2}\mathbb{E}[D+\pi^\star(D)]\nonumber\\
	&\overset{(b)}{\geq}\frac{1}{2}\mathbb{E}[D]\overset{(c)}{\geq}\frac{1}{2}\overline{D}_{\mathsf{lb}}, 
\end{align}
where inequality (a) is obtained by Jensen's inequality $\mathbb{E}\left[(D+\pi^\star(D))^2\right]\geq\mathbb{E}[(D+\pi^\star(D))]^2$; inequality (b) is because $0\leq\pi(D)\leq W_{\mathsf{ub}}$ and the non-negativity of $D$; inequality $(c)$ obtained due to Assumption~\ref{assu:delay}. 

To establish the upper bound of $\gamma^\star$, we consider the constant wait policy $\pi_{\mathsf{const}}$, namely the waiting interval is fixed as a constant $W_k\equiv\frac{1}{f_{\mathsf{max}}}$ for any cycle $k$. According to \eqref{eq:rrobj}, the expected average AoI of policy $\pi_{\mathsf{const}}$ can be computed by:
\begin{align}
\overline{A}_{\pi_{\mathsf{const}}}&=\frac{\mathbb{E}\left[\frac{1}{2}(D+\pi_{\mathsf{const}}(D))^2\right]}{\mathbb{E}\left[D+\pi_{\mathsf{const}}(D)\right]}+\overline{D}\nonumber\\
&\leq \frac{\frac{1}{2}M_{\mathsf{ub}}+\overline{D}\frac{1}{f_{\mathsf{max}}}+\frac{1}{2}\frac{1}{f_{\mathsf{max}}^2}}{\overline{D}_{\mathsf{lb}}+\frac{1}{f_{\mathsf{max}}}}+\overline{D}. 
\end{align}

Notice that policy $\pi_{\mathsf{const}}$ may not be the AoI optimum strategy, i.e., $\overline{A}_{\pi^\star}\leq\overline{A}_{\pi_\mathsf{const}}$. Recall that the optimum ratio is computed by $\gamma^\star=\overline{A}_{\pi^\star}-\overline{D}$, we have:
\begin{equation}
	\gamma^\star\leq\overline{A}_{\pi_{\mathsf{const}}}-\overline{D}\leq\frac{\frac{1}{2}M_{\mathsf{ub}}+\overline{D}_{\mathsf{ub}}\frac{1}{f_{\mathsf{max}}}+\frac{1}{2}\frac{1}{f_{\mathsf{max}}^2}}{\overline{D}_{\mathsf{lb}}+\frac{1}{f_{\mathsf{max}}}}=:\gamma_{\mathsf{ub}}. 
\end{equation}
\end{IEEEproof}
\section{Proof of Theorem~\ref{thm:regret}}\label{pf:regret}
\begin{IEEEproof}First, recall that the ratio $\gamma_k$ in any cycle $k$ is upper bounded by $\gamma_{\mathsf{ub}}$, since the transmission delay is bounded $D\leq B$, the length $L_k$ and reward $Q_k$ in cycle $k$ can be upper bounded by:
    \begin{align}
    	L&\leq D+(\gamma-D)^+\leq B+\gamma_{\mathsf{ub}}=:L_{\mathsf{ub}},\nonumber\\
    	Q&=\frac{1}{2}L^2\leq L_{\mathsf{ub}}^2.
    \end{align}
    
    Let $\overline{L}^\star:=\mathbb{E}[D+\pi^\star(D)]$ and $\overline{Q}^\star:=\mathbb{E}[\frac{1}{2}(D+\pi^\star(D))^2]$ be the expected average cycle length and the expected average reward if the optimum policy $\pi^\star$ is used. We will first provide the following lemmas:
\begin{lemma}\label{lemma:cons-1}
	The expected cycle length $\mathbb{E}[L_{k}|\gamma_k]$ and the expected reward $\mathbb{E}[Q_{k}|\gamma_k]$ received in cycle $k$ satisfies:
	\begin{subequations}
		\begin{align}
		&\mathbb{E}\left[Q_k-\gamma_k L_{k}|\gamma_k\right]\leq(\gamma^\star-\gamma_{ k})\overline{L}^\star,\label{eq:lemmacons-1-eqn1}\\
		&\mathbb{E}\left[Q_k-\gamma^\star L_{k}|\gamma_k\right]\leq-(\gamma^\star-\gamma_{k})\left(\mathbb{E}[L_{ k}|\gamma_k]-\overline{L}^\star\right). \label{eq:lemmacons-1-eqn2}
		\end{align}
	\end{subequations}
\end{lemma}

\begin{lemma}{\label{lemma:cons-2new}}
Recall from equation \eqref{eq:aoidecomp}, the cumulative AoI in cycle $k$ is $X_k=Q_k+L_{k-1}D_k$. The cumulative AoI up to the end of cycle $K$, i.e.,  $\mathbb{E}\left[\int_{0}^{S_{K+1}}A(t)\text{d}t\right]=\mathbb{E}\left[\sum_{k=1}^KX_k\right]$, satisfies the following inequality: 
	\begin{align}
	\mathbb{E}\left[\sum_{k=1}^K(X_{k}-(\gamma^\star+\overline{D}) L_{k})\right]\leq\mathbb{E}\left[\sum_{k=1}^K(\gamma^\star-\gamma_{k})^2\right].\label{eq:lemma-2}
	\end{align}
\end{lemma}

Proofs for Lemma~\ref{lemma:cons-1} and \ref{lemma:cons-2new} are provided in Appendix \ref{pf:cons-1} and \ref{pf:lemma:cons-1}. Through \eqref{eq:lemma-2}, the average cost deviation can be upper bounded by:
 \begin{align}
&\tilde{A}_K-\overline{A}_{\pi^\star}\nonumber\\
=&\frac{\mathbb{E}\left[\sum_{k=1}^KX_k\right]}{\mathbb{E}\left[\sum_{k=1}^KL_{k}\right]}-(\gamma^\star+\overline{D})\nonumber\\
=&\frac{\mathbb{E}\left[\sum_{k=1}^K(X_{k}-(\gamma^\star+\overline{D}) L_{k})\right]}{\mathbb{E}\left[\sum_{k=1}^KL_{k}\right]}\nonumber\\
\leq& \frac{\mathbb{E}\left[\sum_{k=1}^K(\gamma^\star-\gamma_{k})^2\right]}{\mathbb{E}\left[\sum_{k=1}^KL_{k}\right]}.\label{eq:atildeub}
\end{align} 

We then prove inequalities in Theorem~\ref{thm:regret} as follows: 

\subsection{Proof of \eqref{eq:theorem4-gammadiff}}
For simplicity, denote \begin{align}
z_{k+1}:=\gamma_{k}+\eta_k(Q_k-\gamma_k L_{ k}). \label{eq:zdef}
\end{align}

Since $\gamma_{k+1}=[z_{ k+1}]_{\gamma_{\mathsf{lb}}}^{\gamma_{\mathsf{ub}}}$ and $\gamma^\star\in[\gamma_{\mathsf{lb}}, \gamma_{\mathsf{ub}}]$, we can bound the stepsize deviation $(\gamma_{k+1}-\gamma^\star)^2$ using $(z_{k+1}-\gamma^\star)^2$:
\begin{equation}
(\gamma_{k+1}-\gamma^{\star})^2=([z_{ k+1}]_{\gamma_{\mathsf{lb}}}^{\gamma_{\mathsf{ub}}}-[\gamma^{\star}]_{\gamma_{\mathsf{lb}}}^{\gamma_{\mathsf{ub}}})\leq(z_{k+1}-\gamma^{\star})^2. \label{eq:lemm4-gammtoz}
\end{equation}

We proceed to upper bound $(z_{k+1}-\gamma^\star)^2$ as follows:
\begin{align}
&\frac{1}{2}(z_{k+1}-\gamma^{\star})^2\nonumber\\
\overset{(a)}{=}&\frac{1}{2}\left(\gamma_{k}-\gamma^{\star}+\eta_k\left(Q_k-\gamma_{ k}L_{ k}\right)\right)^2\nonumber\\
=&\frac{1}{2}(\gamma_{k}-\gamma^{ \star})^2+\frac{1}{2}\eta_{k}^2\left(Q_k-\gamma_{k} L_{k}\right)^2\nonumber\\
&+\eta_k(\gamma_{k}-\gamma^{\star})\left(Q_k-\gamma_{k}L_{k}\right)\nonumber\\
\overset{(b)}{\leq}&\frac{1}{2}(\gamma_{k}-\gamma^{\star})^2+\frac{1}{2}\eta_k^2L_{\mathsf{ub}}^4+\eta_k(\gamma_{k}-\gamma^{\star})\left(Q_k-\gamma_{k} L_{k}\right),
\label{eq:step-dminimish1}
\end{align}
where equality (a) is obtained from the definition of $z_k$ in \eqref{eq:zdef}; inequality (b) is obtained because $Q_k=\frac{1}{2}L_k^2\leq L_{\mathsf{ub}}^2$ and $\gamma_kL_k\leq L_{\mathsf{ub}}^2$. Then, taking the conditional expectation on both sides of~\eqref{eq:step-dminimish1}, we have:
\begin{align}
&\frac{1}{2}\mathbb{E}\left[(z_{ k+1}-\gamma^{\star})^2|\gamma_k\right]\nonumber\\
\leq&\frac{1}{2}(\gamma_{k}-\gamma^{\star})^2+\frac{1}{2}\eta_k^2L_{\mathsf{ub}}^4\nonumber\\
&+\eta_k(\gamma_{k}-\gamma^{\star})\mathbb{E}\left[Q_k-\gamma_{k}L_{ k}|\gamma_k\right].\label{eq:lemm4-drift}
\end{align}

We then proceed to bound the last term in~\eqref{eq:lemm4-drift}, i.e., 
\begin{equation}
    (\gamma_{k}-\gamma^{\star})\mathbb{E}\left[Q_k-\gamma_{k}L_{ k}|\gamma_k\right]\label{eq:lasterm}
\end{equation}
\begin{itemize}
	\item If the current $\gamma_{ k}-\gamma^{\star}\geq 0$, by plugging \eqref{eq:lemmacons-1-eqn1} into \eqref{eq:lasterm}, we have:
	\begin{align}
	&(\gamma_{k}-\gamma^{\star})\mathbb{E}[Q_k-\gamma_{k}L_{ k}|\gamma_k]\nonumber\\
	\leq&-(\gamma_{k}-\gamma^{\star})^2\overline{L}^{\star}\leq-(\gamma_k-\gamma^\star)^2\overline{D},
	\label{eq:lemm4-ub3-1}
	\end{align}
	where the last inequality is obtained because $\overline{L}^\star\geq\overline{D}$. 
	
	\item  If the current $\gamma_{ k}-\gamma^{\star}\leq 0$, we can upper the last term in inequality \eqref{eq:lemm4-drift} as follows:
	\begin{align}
	&(\gamma_{k}-\gamma^{\star})\mathbb{E}[Q_k-\gamma_{k}L_{ k}|\gamma_k]\nonumber\\
	=&(\gamma_{k}-\gamma^{\star})\mathbb{E}[Q_k-\gamma^{\star}L_{ k}|\gamma_k]\nonumber\\
	&-(\gamma_{k}-\gamma^{\star})^2\mathbb{E}[L_{ k}|\gamma_k]\nonumber\\
	\overset{(c)}{\leq}&(\gamma_{k}-\gamma^{\star})(\overline{Q}^{\star}-\gamma^{\star}\overline{L}^{\star})-(\gamma_{k}-\gamma^{\star})^2\mathbb{E}[L_{ k}|\gamma_k]\nonumber\\
	=&-(\gamma_{k}-\gamma^{\star})^2\mathbb{E}[L_{ k}|\gamma_k]\nonumber\\
	\overset{(d)}{\leq}&-(\gamma_{k}-\gamma^{\star})^2\overline{D},\label{eq:lemm4-ub3-2}
	\end{align}
	where inequality (c) is because $\mathbb{E}[Q_k-\gamma^\star L_k|\gamma_k]\geq\overline{Q}^\star-\gamma^\star\overline{L}^\star=0$ and inequality (d) is because $\mathbb{E}[L_k|\gamma_k]\geq\overline{D}$. 
\end{itemize}
Plugging \eqref{eq:lemm4-ub3-1} and \eqref{eq:lemm4-ub3-2} into \eqref{eq:lemm4-drift}, then taking the expectation with respect to $\gamma_k$ yields:
\begin{align}
&\frac{1}{2}\mathbb{E}\left[(z_{k+1}-\gamma^{\star})^2|\gamma_k\right]\nonumber\\
=&\left(\frac{1}{2}-\eta_k\overline{D}\right)\mathbb{E}\left[(\gamma_k-\gamma^\star)^2\right]+\frac{1}{2}\eta_k^2L_{\mathsf{ub}}^4\nonumber\\
\leq&\left(\frac{1}{2}-\eta_k\overline{D}_{\mathsf{lb}}\right)\mathbb{E}\left[(\gamma_k-\gamma^\star)^2\right]+\frac{1}{2}\eta_k^2L_{\mathsf{ub}}^4.\label{eq:zkub}
\end{align}

By taking the expectation of inequality~\eqref{eq:zkub} with respect to ratio $\gamma_k$ and plugging in it into \eqref{eq:lemm4-gammtoz}, we can upper bound $\mathbb{E}[(\gamma_{k+1}-\gamma^\star)^2]$ by:
\begin{align}
&\frac{1}{2}\mathbb{E}\left[(\gamma_{k+1}-\gamma^{\star})^2\right]\nonumber\\
\leq&\frac{1}{2}\left(1-2\eta_k\overline{D}_{\mathsf{lb}}\right)\mathbb{E}\left[(\gamma_{ k}-\gamma^{ \star})^2\right]+\frac{1}{2}\eta_k^2L_{\mathsf{ub}}^4. \label{eq:gammakevolve}
\end{align}

Next, by choosing stepsizes $\eta_1=\frac{1}{2\overline{D}_{\mathsf{lb}}}$ and $\eta_k=\frac{1}{(k+2)\overline{D}_{\mathsf{lb}}}, \forall k>1$, we can then show by induction that 
\begin{equation}
\frac{1}{2}\mathbb{E}[(\gamma_{k}-\gamma^{ \star})^2]\leq\frac{1}{2k}\frac{L_{\mathsf{ub}}^4}{\overline{D}_{\mathsf{lb}}^2}. \label{eq:theorm4-step-diminish}
\end{equation}

The proof is as follows:

\begin{itemize}
	\item When $k=2$, plugging the stepsize $\eta_1=\frac{1}{2\overline{D}_{\text{lb}}}$ into \eqref{eq:gammakevolve} yields:
	\[\frac{1}{2}\mathbb{E}[(\gamma_2-\gamma^\star)^2]\leq\frac{1}{8}\frac{L_{\mathsf{ub}}^4}{\overline{D}_{\mathsf{lb}}^2}\leq\frac{1}{4}\frac{L_{\mathsf{ub}}^4}{\overline{D}_{\mathsf{lb}}^2}.\]
	
	\item When $k>2$, assuming that $\frac{1}{2}\mathbb{E}[(\gamma_{k}-\gamma^\star)^2]\leq\frac{1}{2k}\frac{L_{\mathsf{ub}}^4}{\overline{D}_{\mathsf{lb}}^2}$, recall that the stepsize $\eta_k=\frac{1}{(k+2)\overline{D}_{\mathsf{lb}}}$, we have
	\begin{align}
	&\frac{1}{2}\mathbb{E}\left[(\gamma_{k+1}-\gamma^\star)^2\right]\nonumber\\
	\leq&\left(\frac{1}{2}-\eta_k\overline{D}_{\mathsf{lb}}\right)\mathbb{E}\left[(\gamma_k-\gamma^\star)^2\right]+\frac{1}{2}\eta_k^2L_{\mathsf{ub}}^4\nonumber\\
	\leq&\left(1-\frac{2}{k+2}\right)\frac{1}{2k}\frac{L_{\mathsf{ub}}^4}{\overline{D}_{\mathsf{lb}}^2}+\frac{1}{2}\frac{1}{(k+2)^2}\frac{L_{\mathsf{ub}}^4}{\overline{D}_{\mathsf{lb}}^2}\nonumber\\
	=&\frac{1}{2}\left(\frac{1}{k+2}+\frac{1}{(k+2)^2}\right)\frac{L_{\mathsf{ub}}^4}{\overline{D}_{\mathsf{lb}}^2}\nonumber\\
	=&\frac{1}{2}\frac{k+3}{(k+2)^2}\frac{L_{\mathsf{ub}}^4}{\overline{D}_{\mathsf{lb}}^2}\nonumber\\
	\overset{(f)}{\leq}&\frac{1}{2}\frac{1}{(k+1)}\frac{L_{\mathsf{ub}}^4}{\overline{D}_{\mathsf{lb}}^2},
	\end{align}
	where inequality (f) is obtained because $(k+1)(k+3)\leq (k+2)^2$. 
\end{itemize}
\subsection{Proof of \eqref{eq:theorem4-conclusion1}} Summing up the inequality \eqref{eq:theorem4-gammadiff} from cycle $k=1$ to $K$ we have:
\begin{align}
&\mathbb{E}\left[\sum_{k=1}^K(\gamma^\star-\gamma_{k})^2\right]\nonumber\\
\leq&\frac{L_{\mathsf{ub}}^4}{\overline{D}_{\mathsf{lb}}^2}\left(\sum_{k=1}^K\frac{1}{k}\right)\nonumber\\
\overset{(a)}{\leq}&\frac{L_{\mathsf{ub}}^4}{\overline{D}_{\mathsf{lb}}^2}\left( 1+\int_{k=1}^{K}\frac{1}{k}\text{d}k\right)\nonumber\\
\overset{(b)}{=}&\frac{L_{\mathsf{ub}}^4}{\overline{D}_{\mathsf{lb}}^2}\left(1+\ln K\right),\label{eq:perstepub}
\end{align}
where inequality $(a)$ is obtained because $\frac{1}{k}\leq\int_{k'=k-1}^{k}\frac{1}{k'}\text{d}k', \forall k>1$ and equality $(b)$ is obtained because $\int_{a}^b\frac{1}{x}\text{d}x=\ln b-\ln a$. 
Plugging inequality \eqref{eq:perstepub} into \eqref{eq:atildeub} we have:
\begin{align}
\tilde{A}_K-\overline{A}_{\pi^\star}=&\frac{\mathbb{E}\left[\sum_{k=1}^K(\gamma^\star-\gamma_{k})^2\right]}{\mathbb{E}\left[\sum_{k=1}^KL_{k}\right]}\nonumber\\
\leq&\frac{L_{\mathsf{ub}}^4}{\overline{D}_{\mathsf{lb}}^2}\left(1+\ln K\right)\frac{1}{\mathbb{E}\left[\sum_{k=1}^KL_k\right]}\nonumber\\
\overset{(c)}{\leq}&\frac{L_{\mathsf{ub}}^4}{\overline{D}\overline{D}_{\mathsf{lb}}^2}\times \frac{1+\ln K}{K},
\end{align}
where inequality $(c)$ is because $\mathbb{E}\left[\sum_{k=1}^KL_k\right]\geq\mathbb{E}\left[\sum_{k=1}^KD_k\right]= K\overline{D}$. This finishes the proof of \eqref{eq:theorem4-conclusion1}.
\subsection{Proof of \eqref{eq:theorem4-conclusion2}}
Recall that the expected time average AoI using stationary policy $\pi_K$ with ratio $\gamma_K$ can be computed by \[\overline{A}_{\pi_K}=\frac{\mathbb{E}[\frac{1}{2}((\gamma_K-D)^++D)^2]}{\mathbb{E}[(\gamma_K-D)^++D]}+\overline{D}.\]

Since $\pi^\star$ is the optimum stationary policy that achieves the smallest AoI, therefore for any stationary policy $\pi_K$, we have $\overline{A}_{\pi_K}\geq\overline{A}_{\pi^\star}$ and the optimality gap can be upper bounded by:
\begin{align}
	&\overline{A}_{\pi_K}-\overline{A}_{\pi^\star}\nonumber\\
	=&\frac{\mathbb{E}\left[\frac{1}{2}((\gamma_K-D)^++D)^2\right]}{\mathbb{E}\left[(\gamma_K-D)^++D\right]}-\gamma^\star\nonumber\\
	=&\frac{\mathbb{E}\left[\frac{1}{2}((\gamma_K-D)^++D)^2-\gamma_K((\gamma_K-D)^++D)\right]}{\mathbb{E}[(\gamma_K-D)^++D]}\nonumber\\
	&+(\gamma_K-\gamma^\star)\nonumber\\
	\overset{(d)}{=}&\frac{\mathbb{E}\left[Q_K-\gamma_KL_K\right]}{\mathbb{E}[L_K]}+(\gamma_K-\gamma^\star)\nonumber\\
	\overset{(e)}{\leq}&\frac{(\gamma^\star-\gamma_K)\overline{L}^\star}{\mathbb{E}[(\gamma_K-D)^++D]}+(\gamma_K-\gamma^\star)\nonumber\\
	=&(\gamma_K-\gamma^\star)\nonumber\\
 &\times\left(\frac{\mathbb{E}\left[(\gamma_K-D)^++D\right]-\mathbb{E}\left[(\gamma^\star-D)^++D\right]}{\mathbb{E}\left[(\gamma_K-D)^++D\right]}\right)\nonumber\\
	=&\frac{(\gamma_K-\gamma^\star)}{\mathbb{E}[(\gamma_K-D)^++D]}\mathbb{E}[(\gamma_K-D)^+-(\gamma^\star-D)^+]\nonumber\\
	\leq&\frac{1}{\overline{D}}(\gamma_K-\gamma^\star)^2,\label{eq:pf-2}
\end{align}
where equality $(d)$ is by definition that $Q_K=\frac{1}{2}((\gamma_K-D_K)^++D_K)^2$, $L_K=(\gamma_K-D_K)^++D_K$ and the transmission delay $D_K$ is i.i.d.; inequality $(e)$ is obtained by taking the expectation with respect to $\gamma_k$ of inequality  $\mathbb{E}[Q_K-\gamma_KL_K|\gamma_k]\leq(\gamma^\star-\gamma_K)\overline{L}^
\star$ from Lemma~\ref{lemma:cons-1}.

Plugging \eqref{eq:step-dminimish1} into inequality \eqref{eq:pf-2}, we can then complete the proof of Theorem~\ref{thm:regret}: \[\mathbb{E}\left[\overline{A}_{\pi_k}-\overline{A}_{\pi^\star}\right]\leq\frac{L_\mathsf{ub}^4}{\overline{D}\overline{D}_{\mathsf{lb}}^2}\frac{1}{k}. \]
\end{IEEEproof}

\section{Proof of Theorem~\ref{thm:converge}}\label{pf:as}

\subsection{Proof of \eqref{eq:thm1-1}}

The proof is divided into two steps, first we will show that $\{\gamma_k\}$ converges to the limit points of an Ordinary Differential Equation (ODE) with probability 1, and then we will show that the $\gamma^\star$ is the unique stationary point of the ODE. 

Notice that when there is no sampling frequency constraint, $\nu_k\equiv 0$. For each $D\sim\mathbb{P}_D$, define function
\begin{equation}
g(\gamma;D):=\frac{1}{2}((\gamma-D)^++D)^2-\gamma((\gamma-D)^++D),
\end{equation}
and the expectation over $\mathbb{P}_D$ is denoted by:
\begin{equation}
\overline{g}(\gamma):=\mathbb{E}\left[g(\gamma;D)\right]. 
\end{equation}

With function $g$, the update rule in equation~\eqref{eq:robbins-monro-gamma} can be rewritten as follows:
\begin{equation}
	\gamma_{k+1}=\left[\gamma_k+\eta_k Y_k\right]_{\gamma_{\mathsf{lb}}}^{\gamma_{\mathsf{ub}}},
\end{equation}
where $Y_k:=g(\gamma_k;D_k)$. 

Next, we will show that the update step-size $\{\eta_k\}$ and $Y_k$ satisfy the following properties:
\begin{itemize}
	\item[(1.1)] Since $\gamma_k$ is bounded, the second order moment of $Y_k$ is bounded, i.e., 
	\begin{align}
		\mathbb{E}\left[\left|Y_k\right|^2\right]=&\mathbb{E}\left[\left(Q_k-\gamma_kL_k\right)^2\right]\nonumber\\
		\leq&\mathbb{E}\left[\left(\frac{1}{2}((\gamma_k-D_k)^++D_k)^2\right)^2\right]\nonumber\\
  &+\mathbb{E}\left[\gamma_k^2\left((\gamma_k-D_k)^++D_k\right)^2\right]<\infty. 
	\end{align}
	\item[(1.2)] Since $D_k$ appears i.i.d. and $\gamma_k$ is determined by historical $\{Y_i\}_{i\leq k-1}$, we have
	\begin{align}
		&\mathbb{E}\left[Y_k\right]=\mathbb{E}[Y_k|\gamma_1, \{Y_i\}_{i\leq k-1}]\nonumber\\
  =&\mathbb{E}\left[g(\gamma_k, D_k)|\gamma_k\right]=\overline{g}(\gamma_k). 
	\end{align}
	
	\item[(1.3)]  Function $\overline{g}(\cdot)$ is continuous. 
\end{itemize}

Notice that the step-sizes $\{\eta_k\}$ are chosen such that $\sum_{k=1}^\infty\eta_k=\infty$ and $\sum_{k=1}^\infty\eta_k^2<\infty$. The ratio in the $k$-th cycle on sample path $\omega$ is denoted by $\gamma_k(\omega)$, according to \cite[p.126, Theorem 2.1]{Kushner2003}, with probability 1, the limits $\gamma_k(\omega)$ are trajectories of the following ordinary differential equation:
\begin{equation}
	\dot{\gamma}=\overline{g}(\gamma). \label{eq:ode1}
\end{equation}

We will then show that $\gamma^\star$ is the unique stationary point of ODE~\eqref{eq:ode1}. The derivative $\overline{g}(\gamma)$ can be computed by:
\begin{equation}
\overline{g}'(\gamma)=-\gamma\cdot\text{Pr}\left(D\leq\gamma\right),\label{eq:mono}
\end{equation}

Therefore, function $\overline{g}(\gamma)$ is monotonically non-increasing over $\mathbb{R}^+$, and is  monotonically decreasing for $\gamma$ that satisfies $\text{Pr}(\gamma>D)>0$. Therefore, if zero-wait policy is not optimum, i.e., $\text{Pr}(\gamma^\star>0)>0$, then $\overline{g}(\gamma^\star)=0$ and $\gamma^\star$ is the unique solution to the following equation
\begin{equation}
\overline{g}(\gamma)=0. 
\end{equation}

We will then show $\gamma^\star$ is the unique stationary point of ODE~\eqref{eq:ode1} through Lyapunov stability analysis, where the Lyapunov function is denoted by $V(\gamma):=\frac{1}{2}(\gamma-\gamma^\star)$. Then we have:
\begin{equation}
	\dot V(\gamma)=(\gamma-\gamma^\star)\overline{g}(\gamma).
\end{equation}

According to the monotonic characteristic from \eqref{eq:mono}, we have $\dot{V}(\gamma)<0, \forall \gamma\neq\gamma^\star$ and the global stability of $\gamma^\star$ is verified from Lyapunov theorem. Since $\{\gamma_k\}$ almost surely to the limit point of the ODE \eqref{eq:ode1} and $\gamma^\star$ is the unique stationary point of \eqref{eq:ode1}, we conclude that $\gamma_k$ converges to $\gamma^\star$ almost surely.

\subsection{Proof of \eqref{eq:theorem4-as-avgaoi}}

Let $a_k$ be the average AoI up to frame $k$, which can be computed by:
\begin{equation}
a_k:=\frac{\int_{t=0}^{S_{k+1}}A(t)\text{d}t}{S_{k+1}}=\frac{\frac{1}{k}\int_{t=0}^{S_{k+1}}A(t)\text{d}t}{\frac{1}{k}S_{k+1}}. \label{eq:adef}
\end{equation}

To show that sequence $\{a_k\}$ converges to $\overline{A}_{\pi^\star}$ almost surely, we will first show that the denominator in \eqref{eq:adef} is strictly positive with probability 1. Notice that $\frac{1}{k}S_{k+1}$ can be computed by:
\begin{equation}
	\frac{1}{k}S_{k+1}=\frac{1}{k}\sum_{k'=1}^k(D_{k'}+W_{k'})\geq\frac{1}{k}\sum_{k'=1}^kD_{k'}.\label{eq:denom}
\end{equation}

Since the transmission delays $\{D_{k'}\}$ are i.i.d., taking the limit on both sides inequality \eqref{eq:denom}, the law of large number shows:
\begin{equation}
	\liminf_{k\rightarrow\infty}\frac{1}{k}S_{k+1}\geq\liminf\frac{1}{k}\sum_{k'=1}^kD_{k'}\overset{\text{a.s.}}{=}\overline{D}>0. \label{eq:denompositive}
\end{equation}

Inequality \eqref{eq:denompositive} implies, sequence $\frac{1}{k}S_{k+1}$ is strictly larger than a positive constant with probability 1. To prove sequence $\{a_k\}=\frac{\int_{t=0}^{S_{k+1}}A(t)\text{d}t}{S_{k+1}}$ converges to $\overline{A}_{\pi^\star}$, it is equivalent to show that \begin{align}
&\lim_{k\rightarrow\infty}\theta_k\overset{\text{a.s.}}{=}0, \label{eq:theta-as}\\
&\text{ where }\theta_k:=\frac{1}{k}\int_{t=0}^{S_{k+1}}A(t)\text{d}t-\overline{A}_{\pi^\star}\cdot\left(\frac{1}{k}S_{k+1}\right). \nonumber
\end{align}

The proof will proceed in two steps: (i) we will show that with probability 1, $\{\theta_k\}$ converges to the limit points of an ODE; (ii) we will show that $0$ is the unique stationary point of the ODE. The first step is to rewrite the evolution of $\{\theta_k\}$ into a recursive form. Recall that the cumulative AoI in frame $k$ is $\int_{S_k}^{S_{k+1}}A(t)\text{d}t=Q_k+L_{k-1}D_k$ and the optimum AoI $\overline{A}_{\pi^\star}=\gamma^\star+\overline{D}$, $\theta_k$ can be rewritten as follows:
\begin{align}
	\theta_k=&\frac{1}{k}\sum_{k'=1}^k(Q_{k'}+L_{k'-1}D_{k'})\nonumber\\
 &-\left(\gamma^\star+\overline{D}\right)\cdot\left(\frac{1}{k}\sum_{k'=1}^kL_{k'}\right)\nonumber\\
	=&\frac{1}{k}\left((k-1)\theta_{k-1}+Q_k+L_{k-1}D_k-(\gamma^\star+\overline{D})L_{k}\right)\nonumber\\
	=&\theta_{k-1}+\frac{1}{k}\underbrace{\left(Q_k-(\gamma^\star+\overline{D})L_k-\theta_{k-1}+L_{k-1}D_k\right)}_{=:Y_k}\nonumber\\
	=&\theta_{k-1}+\frac{1}{k}\left(\mathbb{E}[Y_k|\mathcal{H}_{k-1}]+(Y_k-\mathbb{E}[Y_k|\mathcal{H}_{k-1}])\right)\label{eq:theta-recurse-initial}
\end{align}

To further simply the evolution of $\theta_k$, we make the following definitions on function $f(\theta, \gamma;d)$:
\begin{equation}
	f(\theta, \gamma;d):=\frac{1}{2}\left((\gamma-d)^++d\right)^2-\gamma\cdot\left((\gamma-d)^++d\right)-\theta.
\end{equation}

Let $f(\theta, \gamma):=\mathbb{E}_D[f(\theta, \gamma;D)]$ be the expectation over $D$. Specifically, denote function $\overline{f}(\theta)$ as the value of $f(\theta, \gamma)$ when $(\gamma=\gamma^\star)$. By definition $\overline{f}(\theta)$ can be simplified as follows:
\begin{align}
	\overline{f}(\theta):=&f(\theta, \gamma^\star)\nonumber\\
	=&\mathbb{E}_D\Big[\frac{1}{2}\left((\gamma^\star-D)^++D\right)^2\nonumber\\
 &-\gamma^\star\cdot\left(\left(\gamma^\star-D\right)^++D\right)\Big]-\theta\nonumber\\
	\overset{(a)}{=}&-\theta,
\end{align}
where equality $(a)$ is because $\mathbb{E}\left[\frac{1}{2}\left((\gamma^\star-D^++D)\right)^2\right]-\gamma^\star\mathbb{E}\left[(\gamma^\star-D)^++D\right]=0$. 

Then given historical transmission $\mathcal{H}_{k-1}$, the conditional expectation of $Y_k$ can be computed by:
\begin{align}
	&\mathbb{E}\left[Y_k|\mathcal{H}_{k-1}\right]\nonumber\\
 =&\mathbb{E}[f(\theta_{k-1}, \gamma_k;D)]-\overline{D}\mathbb{E}[L_k|\gamma_k]+L_{k-1}\overline{D}\nonumber\\
	=&f(\theta_{k-1}, \gamma_k)\nonumber\\
 &-\underbrace{\overline{D}\cdot\left(\mathbb{E}\left[(\gamma_k-D)^++D\right]-\mathbb{E}\left[(\gamma^\star-D)^++D\right]\right)}_{=:\beta_{k, 1}}\nonumber\\
&+\underbrace{\overline{D}\cdot\left((\gamma_{k\!-\!1}\!-\!D_{k\!-\!1})^+\!+\!D_{k\!-\!1})\!-\!\mathbb{E}\left[(\gamma_{k\!-\!1}\!-\!D)^+\!+\!D\right]\right)}_{=:\beta_{k, 2}}\label{eq:deltaybias}\end{align}

Finally, denote $\delta M_k:=Y_k-\mathbb{E}[Y_k|\mathcal{H}_{k-1}]$ and plugging equality~\eqref{eq:deltaybias} into equation~\eqref{eq:theta-recurse-initial}, we have:
\begin{align}
	\theta_k=\theta_{k-1}+\frac{1}{k}\cdot\left(f(\theta_{k-1}, \gamma_k)+\delta M_k-\beta_{k, 1}+\beta_{k, 2})\right),  \label{eq:theta-recurse}
\end{align}

Denote $\epsilon_k:=\frac{1}{k}$, which can be viewed as the step-size for updating $\theta_k$. Term $\beta_{k, 1}$ and $\beta_{k, 2}$ can be viewed as two bias terms. Define $t_0=0$ and the cumulative step-sizes up to cycle $k$ is denoted by $t_k=\sum_{i=0}^{k-1}\epsilon_i$. Therefore, $\ln k\leq t_k\leq 1+\ln (k-1)$. For $t\geq 0$, let $m(t)$ be the unique value such that $t_{m(t)}\leq t<t_{m(t)+1}$. We have \begin{equation}
    m(t)=\lfloor\exp(t)\rfloor.\label{eq:mdef}
\end{equation} 
We then present the following properties about the recursive equation \eqref{eq:theta-recurse}:
\begin{itemize}
	\item[(2.1)] Notice that in each frame $k$, $Q_k, L_k$ are bounded. Therefore, $\theta_k$ is bounded and hence $\sup_k\mathbb{E}[|Y_k|]$ is bounded. 
	
	\item[(2.2)] Function $f(\theta, \gamma)$ is continuous in $\theta$ by definition. 
	
	\item[(2.3)] For each $\theta<\infty$, function $|f(\theta, \gamma)|\leq\mathbb{E}\left[\frac{1}{2}((\gamma-D)^++D)^2\right]+\gamma\mathbb{E}\left[(\gamma-D)^++D\right]<\infty$ is bounded. The difference between $f(\theta, \gamma)$ and $\overline{f}(\theta)$ can be computed by
	\begin{align}
	&\left|f(\theta, \gamma)-\overline{f}(\theta)\right|=\left|\mathbb{E}\left[(\gamma-D)^+-(\gamma^\star-D)^+\right]\right|\nonumber\\
 \leq&|\gamma-\gamma^\star|. \label{eq:fbound}
	\end{align}
	Therefore, for each $k$ we have:
	\begin{align}
		&\text{Pr}\left(\sup_{j\geq k}\left|\sum_{i=k}^j\epsilon_i(f(\theta, \gamma_i)-\overline{f}(\theta))\right|\geq\mu\right)\nonumber\\
		\leq &\frac{\mathbb{E}\left[\sup_{j\geq k}\left|\sum_{i=k}^j\epsilon_i(f(\theta, \gamma_k)-\overline{f}(\theta))\right|\right]}{\mu}\nonumber\\
		\leq &\frac{1}{\mu}\mathbb{E}\left[\sum_{i=k}^\infty\epsilon_i\cdot\left|f(\theta, \gamma_i)-\overline{f}(\theta)\right|\right]\nonumber\\
		\overset{(a)}{\leq}&\frac{1}{\mu}\mathbb{E}\left[\sum_{i=k}^\infty\frac{1}{i^{3/4}}\cdot\left(\frac{1}{i^{1/4}}\cdot\left|\gamma_i-\gamma^\star\right|\right)\right]\nonumber\\
		\overset{(b)}{\leq}&\frac{1}{\mu}\sqrt{\left(\sum_{i=k}^\infty\left(i^{-3/4}\right)^2\right)\cdot\mathbb{E}\left[\sum_{i=k}^\infty\left(i^{-1/2}\cdot\left(\gamma_i-\gamma^\star\right)^2\right)\right]}\nonumber\\
		\overset{(c)}{\leq}&\frac{1}{\mu}\sqrt{\left(\sum_{i=k}^\infty i^{-3/2}\right)\cdot\left(\sum_{i=k}^\infty i^{-1/2}\cdot\frac{L_{\mathsf{ub}}^4}{\overline{D}_{\mathsf{lb}}^2} i^{-1}\right)}\nonumber\\
  \leq&\frac{2}{\mu}\cdot\frac{1}{\sqrt{k-1}}\frac{L_{\mathsf{ub}}^4}{\overline{D}_{\mathsf{lb}}^2}.\label{eq:23-1}
	\end{align}
	where inequality $(a)$ is because \eqref{eq:fbound}; inequality $(b)$ is from Cauchy-Schwarz; inequality $(c)$ is because \eqref{eq:theorem4-gammadiff} from Theorem~\ref{thm:regret}. Taking the limit on both sides of inequality \eqref{eq:23-1}, and recall $m(k)=\lfloor\exp(k)\rfloor$ from equation~\eqref{eq:mdef}, we have:
	\begin{align}
		&\lim_{k\rightarrow\infty}\text{Pr}\left(\sup_{j\geq m(k)}\left|\sum_{i=m(k)}^j\epsilon_i\cdot(g(\theta, \gamma_i)-\overline{g}(\theta))\right|\geq \mu\right)\nonumber\\
  \leq&\lim_{k\rightarrow\infty} \frac{2}{\mu}\cdot\frac{1}{\sqrt{\exp(k)-1}}=0. 
	\end{align}
	\item[(2.4)] Given historical transmission $\mathcal{H}_{k-1}$, the difference $\delta M_k$ only depends on $D_k$ and has mean zero. Since $\gamma_k$ is upper bounded in each frame and the delay $D_k$ is second order bounded, the expectation $Q_k, L_k$ are both upper bounded and the difference sequence $\delta M_k$ is second order bounded. Therefore sequence $M_k:=\sum_{k'=1}^k\epsilon_{k'}\delta M_{k'}$ is also a martingale sequence. According to \cite[Chapter 5, Eq.~(2.6)]{Kushner2003}, for each $\mu>0$, we have
	\begin{align}
		&\lim_{k\rightarrow\infty}\text{Pr}\left(\sup_{j\geq k}\left|\sum_{i=k}^j \epsilon_i\delta M_i\right|\geq\mu\right)\nonumber\\
  =&\lim_{k\rightarrow\infty}\text{Pr}\left(\sup_{j\geq k}\left|M_j-M_k\right|\geq\mu\right)=0. 
	\end{align}
	\item[(2.5)] $\beta_{k, 1}$ and $\beta_{k, 2}$ can be viewed as two bias terms in the recursive form. Next we will show:
	\begin{equation}
		\lim_{k\rightarrow\infty}\text{Pr}\left(\sup_{j\geq k}\left|\sum_{i=k}^j\epsilon_i(\beta_{i, 1}+\beta_{i, 2})\right|\geq\mu\right)=0. \label{eq:2.5}
	\end{equation}
	
	The proof is as follows: through the union bound we have
	\begin{align}
		&\lim_{k\rightarrow\infty}\text{Pr}\left(\sup_{j\geq k}\left|\sum_{i=k}^j\epsilon_i (\beta_{i, 1}+\beta_{i, 2})\right|\geq\mu\right)\nonumber\\
		\leq&\lim_{k\rightarrow\infty}\text{Pr}\left(\sup_{j\geq k}\left|\sum_{i=k}^j\epsilon_i\beta_{i, 1}\right|\geq\mu/2  \right)\nonumber\\
  &+\lim_{k\rightarrow\infty}\text{Pr}\left(\sup_{j\geq k}\left|\sum_{i=k}^j\epsilon_i\beta_{i, 2}\right|\geq \mu/2\right). \label{eq:biasterm}
	\end{align}
	
	For given $k$, we can upper bound the first term in inequality~\eqref{eq:biasterm} as follows:
	\begin{align}
		&\text{Pr}\left(\sup_{j\geq k}\left|\sum_{i=k}^j\epsilon_i\beta_{i, 1}\right|\geq\mu/2\right)\nonumber\\
		\overset{(d)}{\leq}&\frac{\mathbb{E}\left[\sup_{j\geq k}\left|\sum_{i=k}^j\epsilon_i\beta_{i, 1}\right|\right]}{\mu/2}\nonumber\\
		\leq&\frac{2}{\mu}\mathbb{E}\left[\sum_{i=k}^\infty\frac{1}{i}|\beta_{i, 1}|\right]\nonumber\\
		\overset{(e)}{\leq}&\frac{2}{\mu}\sqrt{\left(\sum_{i=k}^\infty\left(\frac{1}{i^{3/4}}\right)^2\right)\cdot\mathbb{E}\left[\sum_{i=k}^\infty\left(\frac{1}{i^{1/4}}\beta_{i, 1}\right)^2\right]}\nonumber\\
		\overset{(f)}\leq&\frac{2}{\mu}\sqrt{\left(\sum_{i=k}^\infty i^{-3/2}\right)\cdot\mathbb{E}\left[\sum_{i=k}^\infty i^{-1/2}(\gamma_i-\gamma^\star)^2\right]}\nonumber\\
		\leq&\frac{2}{\mu}\sqrt{\left(\sum_{i=k}^\infty i^{-3/2}\right)\cdot\left(\sum_{i=k}i^{-3/2}\right)\frac{L_{\mathsf{ub}}^4}{\overline{D}_{\mathsf{lb}}^2}}\nonumber\\
  \leq&\frac{4L_{\mathsf{ub}}^2}{\overline{D}_{\mathsf{lb}}}\frac{1}{\sqrt{k}}. \label{eq:bias-1}
	\end{align}
	where inequality $(d)$ is from Markov inequality; inequality $(e)$ is from Cauchy-Schwarz; inequality $(f)$ comes from \eqref{eq:theorem4-gammadiff} in Theorem~\ref{thm:regret}. 
	Taking the limit with respect to $k$ on both sides of inequality~\eqref{eq:bias-1}, we have:
	\begin{equation}
		\lim_{k\rightarrow\infty}\text{Pr}\left(\sup_{j\geq k}\left|\sum_{i=m(k)}^j\epsilon_i\beta_{i, 1}\right|\geq\mu/2\right)=0. 
	\end{equation}
	
	Notice that the second part $\beta_{k, 2}$ is predicable given historical transmission $\mathcal{H}_{k-1}$. It is also a martingale sequence given $\mathcal{H}_{k-2}$. Therefore, $b_k:=\sum_{k'=1}^k\epsilon_k\beta_{k, 2}$ is also a martingale sequence. Through \cite[Chapter 5, Eq.~(2.6)]{Kushner2003} we can obtain:
	\begin{align}
		&\lim_{k\rightarrow\infty}\text{Pr}\left(\sup_{j\geq k}\left|\sum_{i=k}^j\epsilon_i\beta_{i, 2}\right|\geq \mu/2\right)\nonumber\\
  =&\lim_{k\rightarrow\infty}\text{Pr}\left(\sup_{j\geq k}|b_j-b_k|\geq \mu/2\right)=0. \label{eq:bias2}
	\end{align}
	
	Plugging \eqref{eq:bias-1} and \eqref{eq:bias2} into \eqref{eq:biasterm} verifies \eqref{eq:2.5}. 
	\item[(2.6)] Function $f$ is uniformly bounded for $\theta\in\left[0, 2L_{\mathsf{ub}}^2\right], \gamma\in[\gamma_{\mathsf{lb}}, \gamma_{\mathsf{ub}}]$. 
	
	\item[(2.7)] For each $\gamma$ we have:
	\begin{equation}
		\left|f(\theta_1, \gamma)-f(\theta_2, \gamma)\right|=|\theta_1-\theta_2|,
	\end{equation}
	and $\lim_{\theta\rightarrow\infty}|\theta|=0$. 
	
	\item[(2.8)] Sequence $\frac{1}{k}$ satisfies $\sum_{k'=1}^\infty \frac{1}{k'}=\infty$. 
\end{itemize}

Therefore, according to \cite[p.166, Theorem 1.1]{Kushner2003}\footnote{As mentioned on \cite[p. 166, Eq.~(1.10)]{Kushner2003}, assumption (A1.6) in \cite[p. 165]{Kushner2003} becomes: function $g$ is uniformly bounded, \cite[p. 166, Theorem 1.1]{Kushner2003} is still true.}, with probability 1, sequence $\theta_k$ converges to the limit point of the following ODE:
\begin{equation}
	\dot{\theta}=\overline{f}(\theta)=-\theta. \label{eq:ode-2}
\end{equation}

Notice that $\theta=0$ is the unique stationary point of the ODE \eqref{eq:ode-2}. Therefore,
\begin{align}
	&\lim_{k\rightarrow\infty}\theta_k\!=\!\lim_{k\rightarrow\infty}\frac{1}{k}\left(\int_{t=0}^{S_{k+1}}A(t)\text{d}t\!-\!(\gamma^\star\!+\!\overline{D})S_{k+1}\right)=0, \nonumber\\
 &\text{ w.p.1}.\label{eq:thetaas}
\end{align}

Finally, plugging \eqref{eq:thetaas} into ~\eqref{eq:theta-as} implies:
\begin{equation}
	\lim_{k\rightarrow\infty}\frac{\int_{t=0}^{S_{k+1}}A(t)\text{d}t}{S_{k+1}}\overset{\text{a.s.}}{=}\gamma^\star+\overline{D}=\overline{A}_{\pi^\star}. 
\end{equation}
\section{Proof of Lemma \ref{lemma:cons-1}}\label{pf:cons-1}
\begin{IEEEproof}
	Notice that in each cycle $k$, the waiting time $W_k$ is chosen to minimize the objective function \eqref{eq:consopt}, therefore we have:
	\begin{align}
	&\mathbb{E}\left[Q_k-\gamma_{k}L_{k}|\gamma_k\right]
	\overset{(a)}{\leq}(\overline{Q}^\star-\gamma_k\overline{L}^{\star})\nonumber\\
	\overset{}{=}&(\overline{Q}^\star-\gamma^\star\overline{L}^{\star})+(\gamma^{\star}-\gamma_k)\overline{L}^\star
	\overset{(b)}{=}(\gamma^{\star}-\gamma_k)\overline{L}^\star,\label{eq:lemm7-prev}
	\end{align}
	where equality $(a)$ is because policy $\pi_{k}$ used in cycle $k$ minimizes the Lagrange function. Equality $(b)$ is obtained because on the stationary point $\gamma^\star$ we have $\overline{Q}^{\star}=\gamma^{ \star}\overline{L}^{\star}$. This verifies the first inequality in Lemma~\ref{lemma:cons-1}.
	
	Then, adding $(\gamma_{ k}-\gamma^{\star})\mathbb{E}[L_{ k}|\gamma_k]$ to both sides of \eqref{eq:lemm7-prev} leads to:
	\begin{align}
	&\mathbb{E}\left[Q_k-\gamma^{\star}L_{ k}|\gamma_k\right]\leq(\gamma_{ k}-\gamma^{\star})\mathbb{E}\left[L_{ k}-\overline{L}^{\star}|\gamma_k\right].
	\end{align}
	which verifies the second inequality. \end{IEEEproof}

\section{Proof of Lemma~\ref{lemma:cons-2new}}\label{pf:lemma:cons-1}
\begin{IEEEproof}To find the upper bound of $\mathbb{E}\left[\sum_{k'=1}^k((Q_{k'}+L_{k'-1}D_{k'})-(\gamma^\star+\overline{D})L_{k'})\right]$, first we add $\mathbb{E}[L_{k-1}D_{ k}|\gamma_k]$ on both sides on \eqref{eq:lemmacons-1-eqn2} and obtain:
\begin{align}
    &\mathbb{E}[(Q_k+L_{k-1}D_k)-\gamma^\star L_k|\gamma_k]\nonumber\\
    \leq&-(\gamma^\star-\gamma_k)\left(\mathbb{E}[L_k|\gamma_k]-\overline{L}^\star\right)+\mathbb{E}[L_{k-1}D_k|\gamma_k].\label{eq:addboth}
\end{align}

Next, we can proceed to simplify \eqref{eq:addboth} by:
    \begin{align}
        &\mathbb{E}[(Q_{k}+L_{k-1}D_k)-\gamma^{\star}L_{ k}|\gamma_k]\nonumber\\
        \overset{(a)}{\leq}&-(\gamma^{\star}-\gamma_{ k})\left(\mathbb{E}[L_{ k}|\gamma_k]-\overline{L}^{\star}\right)+L_{k-1}\overline{D}\nonumber\\
        \overset{(b)}{\leq}&(\gamma^{\star}-\gamma_{ k})^2+L_{k-1}\overline{D},\label{eq:lemm4-2}
    \end{align}
    where inequality (a) is because $D_k$ is independent of $L_{k-1}$ and thus $\mathbb{E}[L_{k-1}D_k|\gamma_k]=\mathbb{E}[L_{k-1}]\overline{D}$; inequality (b) is because \begin{align}\mathbb{E}[L_k-\overline{L}^\star|\gamma_k]&=\mathbb{E}\left[(\gamma_k-D)^+-(\gamma^\star-D)^+\right]\nonumber\\
    &\leq|\gamma_k-\gamma^\star|. 
    \end{align}
    
    Summing up inequality \eqref{eq:lemm4-2} from cycle $k=1$ to $K$ and take the expectation with respect to $\gamma_k$, we have:
    \begin{align}
    &\mathbb{E}\left[\sum_{k=1}^K((Q_{k}+L_{k-1}D_{k})-\gamma^{\star}L_{k})\right]\nonumber\\
    \leq&\mathbb{E}\left[\sum_{k=1}^K(\gamma^{\star}-\gamma_{ k})^2\right]-\mathbb{E}\left[\sum_{k=1}^KL_{k}\right]\overline{D}. \label{eq:lemm2-final}
    \end{align}
    
    Deducting $\mathbb{E}\left[\sum_{k=1}^{K}L_k\right]\overline{D}+\mathbb{E}[L_K]\gamma^\star$ on both sides of inequality \eqref{eq:lemm2-final} yields:
    \begin{align}
    	&\mathbb{E}\left[\sum_{k=1}^K((Q_k+L_{k-1}D_k)-(\gamma^\star+\overline{D})L_k)\right]\nonumber\\
    	\leq&\mathbb{E}\left[\sum_{k=1}^K(\gamma^\star-\gamma_k)^2\right]-\mathbb{E}[L_K]\overline{D}\leq\mathbb{E}\left[\sum_{k=1}^K(\gamma^\star-\gamma_k)^2\right]. 
    \end{align}
    This completes the proof of Lemma~\ref{lemma:cons-2new}. 
\end{IEEEproof}

\section{Proof of Theorem~\ref{thm-converse}}\label{pf:thm-converse}

\subsection{Proof of inequality \eqref{eq:gammahat}}\label{pf:minimax}
\begin{IEEEproof}For each distribution $\mathbb{P}$, the optimum ratio $\gamma_{\mathbb{P}}^\star$ satisfies the following equation:
\begin{equation}
    \frac{1}{2}\mathbb{E}\left[((\gamma_{\mathbb{P}}^\star-D)^++D)^2\right]-\gamma_{\mathbb{P}}^\star\mathbb{E}\left[(\gamma_{\mathbb{P}}^\star-D)^++D\right]=0. \label{eq:gammaeqn}
\end{equation}
The minimax estimation error bound on $\hat{\gamma}$ is established through the Le Cam's two point method \cite{Yu1997,le2012asymptotic}. Let $\mathbb{P}_1$ and $\mathbb{P}_2$ be two probability distributions and denote $\gamma_1:=\gamma_{\mathbb{P}_1}^\star$, $\gamma_2:=\gamma_{\mathbb{P}_2}^\star$ for simplicity. Through Le Cam's inequality, we have:
\begin{equation}
    \inf_{\hat{\gamma}}\sup_{\mathbb{P}}\mathbb{E}[(\hat{\gamma}(\mathcal{H}_k)-\gamma_{\mathbb{P}}^\star)^2]\geq (\gamma_1-\gamma_2)^2\cdot \mathbb{P}_1^{\otimes k}\wedge \mathbb{P}_2^{\otimes k},\label{eq:gamma-lecam}
\end{equation}
where $\mathbb{P}\wedge\mathbb{Q}=\int\min\{ \text{d}\mathbb{P}, \text{d}\mathbb{Q}\}$. 

To use the Le Cam's method, the first step is to find two distribution $\mathbb{P}_1,\mathbb{P}_2$ such that the difference $(\gamma_1-\gamma_2)^2$ is large but $\mathbb{P}_1^{\otimes k}\wedge\mathbb{P}_2^{\otimes k}$ can be lower bounded. We consider $\mathbb{P}_1=\text{Uni}([0, 1])$ be the uniform distribution. When $D\sim\mathbb{P}_1$, equation \eqref{eq:gammaeqn} can be simplified into:
    \begin{align}
        -\frac{1}{6}\gamma_1^3-\frac{1}{2}\gamma_1+\frac{1}{6}=0.
    \end{align}
    
    Since $\gamma$ is a real number, according to the solution of cubic equation, we have:
    \begin{align}
        \gamma_1&=\left(\sqrt[3]{\frac{1}{2}+\sqrt{\frac{5}{4}}}+\sqrt[3]{\frac{1}{2}-\sqrt{\frac{5}{4}}}\right).\label{eq:cubicsolve}
    \end{align}
    
    Recall that $\mathbb{P}_1$ is a uniform distribution, therefore the probability of waiting by using the optimum policy $\pi_{\mathbb{P}_1}^\star$ is:
    \begin{equation}
    	p_{\text{w, uni}}:=\text{Pr}\left(D\leq\gamma_1|D\sim\mathbb{P}_1\right)=\gamma_1. 
    \end{equation}
    
    Distribution $\mathbb{P}_2$ is defined through the density function $p_2(x)=\frac{\mathsf{P}_2(\text{d}x)}{\text{d}x}$:
    \begin{equation}
        p_2(x)=\begin{cases}
        1-c\sqrt{1/k}, &0 \le x \leq \delta/2;\\
        1, &\delta/2< x<1-\delta/2;\\
        1+c\sqrt{1/k}, & 1-\delta/2 \le x \le 1;\\
        0, &\text{otherwise}. 
        \end{cases}\label{eq:p2def}
    \end{equation}
    where $\delta=\min\{1/3, p_{\text{w, uni}}/2\}$ and $c<1/2$ is fixed as a constant. 
    
    Lower bounding $(\gamma_2-\gamma_1)^2$ is divided into two steps: first we will prove $\gamma_2\geq \gamma_1$; then we will obtain the lower bound of $\gamma_2$ through Taylor expansion. For simplicity, let function $h_1(\cdot)$ and $h_2(\cdot)$ be:
    \begin{subequations}
    \begin{align}
        &h_1(\gamma):=\nonumber\\
        &\mathbb{E}_{D\sim\mathbb{P}_1}\left[\frac{1}{2}((\gamma-D)^++D)^2-\gamma\left((\gamma-D)^++D\right)\right], \\
        &h_2(\gamma):=\nonumber\\
        &\mathbb{E}_{D\sim\mathbb{P}_2}\left[\frac{1}{2}((\gamma-D)^++D)^2-\gamma\left((\gamma-D)^++D\right)\right]. 
    \end{align}
    \end{subequations}
    
    Then $\gamma_1$ and $\gamma_2$ satisfy $h_1(\gamma_1)=0$ and $h_2(\gamma_2)=0$. 
    
    \textbf{Step 1: Showing $\gamma_2>\gamma_1$. }The derivative of function $h_2(\gamma)$ can be computed by:
    \begin{equation}
        h_2(\gamma)'=-\mathbb{E}_{D\sim\mathbb{P}_2}\left[(\gamma-D)^++D\right]<0. \label{eq:h2derivative}
    \end{equation}
    Therefore, function $h_2(\gamma)$ is monotonically decreasing. 
    
    We will then show $h_2(\gamma_1)>0$. Since $h_1(\gamma_1)=0$, it is sufficient to show that $h_2(\gamma_1)>h_1(\gamma_1)$. The difference $h_2(\gamma)-h_1(\gamma)$ can be computed as follows:
    \begin{align}
        &h_2(\gamma)-h_1(\gamma)\nonumber\\
        =&\mathbb{E}_{\mathbb{P}_2}\left[\frac{1}{2}\left((\gamma-D)^++D\right)^2-\gamma\left((\gamma-D)^++D\right)\right]\nonumber\\
        &-\mathbb{E}_{\mathbb{P}_1}\left[\frac{1}{2}\left((\gamma-D)^++D\right)^2-\gamma\left((\gamma-D)^++D\right)\right]\nonumber\\
        \overset{(a)}{=}&\int_{1-\delta/2}^1\frac{c}{\sqrt{k}}\left(\frac{1}{2}\max\{\gamma, x\}^2-\gamma\max\{\gamma, x\}\right)\text{d}x\nonumber\\
        &-\int_{0}^{\delta/2}\frac{c}{\sqrt{k}}\left(\frac{1}{2}\max\{\gamma, x\}^2-\gamma\max\{\gamma, x\}\right)\text{d}x\nonumber\\
        =&\int_{0}^{\delta/2}\frac{c}{\sqrt{k}}\Big(\frac{1}{2}\left(\max\{\gamma, x+(1-\delta)\}^2-\max\{\gamma, x\}^2\right)\nonumber\\
        &-\gamma\left(\max\{\gamma, x+(1-\delta)\}-\max\{\gamma, x\}\right)\Big)\text{d}x\nonumber\\
        =&\int_{0}^{\delta/2}\frac{c}{\sqrt{k}}\left(\frac{1}{2}\left(\max\{\gamma, x\!+\!(1\!-\!\delta)\}+\max\{\gamma, x\}\right)\!-\!\gamma\right)\nonumber\\
        &\times\left(\max\{\gamma, x+(1-\delta)\}-\max\{\gamma, x\}\right)\text{d}x\nonumber\\
        \overset{(b)}{\geq}& 0,
    \end{align}
    where equality $(a)$ is because $(\gamma-D)^++D=\max\{\gamma, D\}$; inequality $(b)$ is because $\max\{\gamma, x+(1-\delta)\}-\max\{\gamma, x\}\geq 0$ for $\delta<1$, and $\max\{\gamma, x+(1-\delta)\}+\max\{\gamma, x\}\geq 2\gamma$. Therefore, $h_2(\gamma_1)\geq h_1(\gamma_1)=0$. Since $h_2(\gamma_2)=0$ and function $h_2(\cdot)$ is monotonically decreasing, we have $\gamma_2\geq \gamma_1$. 
    
    \textbf{Step 2: Taylor expansion to lower bound $h_2(\gamma_1)$. }Through Taylor expansion, we have:
    \begin{equation}
        (\gamma_2-\gamma_1)=\frac{h_2(\gamma_2)-h_2(\gamma_1)}{h_2'(\gamma)}=\frac{h_2(\gamma_1)-h_2(\gamma_2)}{-h_2'(\gamma)},\label{eq:taylor}
    \end{equation}
    where $\gamma\in[\gamma_1, \gamma_2]$. To lower bound $(\gamma_2-\gamma_1)$, it is suffice to lower bound $h_2(\gamma_1)-h_2(\gamma_2)$ and upper bound $h_2'(\gamma)$. By Corollary~\ref{coro:gammaub}, since $c\leq 1/2$ and $\delta<1$, we can upper bound $\gamma_2$ by:
    \begin{equation}
        \gamma_2\leq\frac{\frac{1}{2}\mathbb{E}_{\mathbb{P}_2}[D^2]}{\overline{D}}\leq\frac{\frac{1}{2}\left(\frac{1}{3}+\frac{\delta}{2}\times\frac{c}{\sqrt{k}}\right)}{1/2}\leq 1. 
    \end{equation}
    
    Therefore, according to \eqref{eq:h2derivative}, for any $\gamma\in[\gamma_1, \gamma_2]$, the derivative $h_2'(\gamma)$ can be upper bounded by:
    \begin{equation}
        |h_2'(\gamma)|=\mathbb{E}_{\mathbb{P}_2}[(\gamma_2-D)^++D]\leq\gamma_2+\mathbb{E}_{\mathbb{P}_2}[D]\leq\frac{3}{2}. \label{eq:hderivalb}
    \end{equation}
    
    Notice that $h_2(\gamma_2)=0$ and $h_1(\gamma_1)=0$, lower bounding $h_2(\gamma_1)-h_2(\gamma_2)$ is equivalent to lower bounding $h_2(\gamma_1)-h_1(\gamma_1)$, which is as follows:
    \begin{align}
        &h_2(\gamma_1)-h_1(\gamma_1)\nonumber\\
        =&\mathbb{E}_{\mathbb{P}_2}\left[\frac{1}{2}\left((\gamma_1-D)^++D\right)^2\!-\!\gamma_1\left((\gamma_1-D)^++D\right)\right]\nonumber\\
        &\!-\!\mathbb{E}_{\mathbb{P}_1}\left[\frac{1}{2}\left((\gamma_1-D)^++D\right)^2\!-\!\gamma_1\left((\gamma_1-D)^++D\right)\right]\nonumber\\
        =&\frac{c}{\sqrt{k}}\!\times\!\int_0^{\delta/2}\Big(\frac{1}{2}\left(\max\{\gamma_1, x\!+\!(1\!-\!\delta)\}^2\!-\!\max\{\gamma_1, x\}^2\right)\nonumber\\
        &-\gamma_1\left(\max\{\gamma_1, x\!+\!(1\!-\!\delta)\}-\max\{\gamma_1, x\}\right)\Big)\text{d}x\nonumber\\
        =:&\frac{c}{\sqrt{k}}N_1. \label{eq:hlb}
    \end{align}
    
    Plugging \eqref{eq:hlb} and \eqref{eq:hderivalb} into \eqref{eq:taylor}, the lower bound on $(\gamma_2-\gamma_1)$ can be obtained by:
    \begin{equation}
        (\gamma_2-\gamma_1)\geq \frac{2N_1c}{3}\frac{1}{\sqrt{k}}. \label{eq:gammadifflb}
    \end{equation}
    
    Next, we proceed to lower bound $\mathbb{P}_1^{\otimes k}\wedge \mathbb{P}_2^{\otimes k}$. Notice that:
    \begin{equation}
        \mathbb{P}_1^{\otimes k}\wedge \mathbb{P}_2^{\otimes k}=1-\frac{1}{2}|\mathbb{P}_1^{\otimes k}-\mathbb{P}_2^{\otimes k}|_1,\label{eq:wedge-kl}
    \end{equation}
    where $|\mathbb{P}-\mathbb{Q}|_1=\int|\text{d}\mathbb{P}-\text{d}\mathbb{Q}|_1$ is the $\ell_1$ distance between probability distribution $\mathbb{P}$ and $\mathbb{Q}$. To lower bound $\mathbb{P}_1^{\otimes k}\wedge \mathbb{P}_2^{\otimes k}$, it is sufficient to upper bound $|\mathbb{P}_1^{\otimes k}-\mathbb{P}_2^{\otimes k}|_1$ as follows:
	\begin{align}
	&\frac{1}{2}\left|\mathbb{P}_1^{\otimes k}-\mathbb{P}_2^{\otimes k}\right|_1\nonumber\\
	\overset{(c)}{\leq}&\sqrt{\frac{1}{2}D_{\mathsf{KL}}(\mathbb{P}_2^{\otimes k}||\mathbb{P}_1^{\otimes k})}\nonumber\\
	=&\sqrt{\frac{1}{2}kD_{\mathsf{KL}}(\mathbb{P}_2||\mathbb{P}_1)}\nonumber\\
	\overset{(d)}{\leq}&\sqrt{\frac{1}{2}k\int_0^1\left(p_2(x)\!-\!1\!+\!\frac{1}{\min\{p_2(x), 1\}}(p_2(x)\!-\!1)^2\right)\text{d}x}\nonumber\\
	\overset{(e)}{\leq}&\sqrt{\frac{1}{2}k\frac{1}{\inf_{0\leq d\leq 1} p_2(d)}\int_0^1(p_2(x)-1)^2\text{d}x}\nonumber\\
	\leq&\sqrt{\frac{1}{2}k\frac{1}{1-c\sqrt{1/k}}\delta\frac{c^2}{k}}\leq\sqrt{\delta c^2},\label{eq:klub-last}
	\end{align}
	where inequality $(c)$ is from Pinsker's inequality; where inequality $(d)$ is because the density function $p_1(x)=1$ for uniform distribution, therefore $D_{\mathsf{KL}}(\mathbb{P}_2||\mathbb{P}_1)=\int_0^1p_2(x)\ln p_2(x)\text{d}x$, where $p_2(x)$ is the density function defined in \eqref{eq:p2def}; inequality $(e)$ is because function $g(t):=(t\ln t)$ is convex, its derivative $g(t)''=1/t$, therefore, through Taylor expansion we have $g(t)\leq g(1)+(t-1)+\frac{1}{2}\frac{1}{\min\{t, 1\}}(t-1)^2=(t-1)+\frac{1}{2}\frac{1}{\min\{t, 1\}}(t-1)^2$.  By choosing $c=1/2$ and recall that $\delta<1$, inequality \eqref{eq:klub-last} can be upper bounded by:
	\begin{equation}
	\frac{1}{2}|\mathbb{P}_1^{\otimes k}-\mathbb{P}_2^{\otimes k}|_1\leq\frac{1}{2}, \label{eq:klub}
	\end{equation}
	
	Plugging \eqref{eq:klub} into \eqref{eq:wedge-kl}, we have:
	\begin{equation}
	    \mathbb{P}_1^{\otimes k}\wedge \mathbb{P}_2^{\otimes k}\geq 1/2. \label{eq:wedgelb}
	\end{equation}
    
   Finally, plugging \eqref{eq:wedgelb} and \eqref{eq:gammadifflb} into the Le Cam's inequality \eqref{eq:gamma-lecam} yields:
   \begin{equation}
       \inf_{\hat{\gamma}}\sup_{\mathbb{P}}\mathbb{E}\left[(\hat{\gamma}(\mathcal{H}_k)-\gamma_{\mathbb{P}}^\star)^2\right]\geq\frac{2N_1^2c^2}{9}\cdot\frac{1}{k},
   \end{equation}
   which verifies \eqref{eq:gammahat}. 

\subsection{Proof of inequality \eqref{eq:avgAoIconverse} }
    We begin the proof of Theorem~\ref{thm:converge} by introducing the following Lemma:
    \begin{lemma}\label{lemma:converse-1}
    Suppose $\gamma^\star$ is the optimum threshold policy $\pi^\star$ selects and let $p_w:=\text{Pr}(D\leq \gamma^\star)$ be the probability of waiting to before taking the next sample. For any stationary policy $\pi$, denote $q_\pi:=\mathbb{E}\left[\frac{1}{2}(D+\pi(D))^2\right]$ and $l_\pi:=\mathbb{E}[D+\pi(D)]$ be the expected average reward and length of each cycle, which satisfy the following inequality:
    \begin{equation}q_\pi\geq\gamma^\star l_\pi+\frac{1}{2}p_w\left(l_\pi-\overline{L}^\star\right)^2.\label{eq:lemma-4}
    \end{equation}
    \end{lemma}
    
Inequality~\eqref{eq:lemma-4} implies, for any causal policy $\pi$, the expected reward and frame length satisfy:
    \begin{align}
        \mathbb{E}\left[Q_k|\mathcal{H}_{k-1}\right]\geq&\gamma^\star\mathbb{E}\left[L_k|\mathcal{H}_{k-1}\right]\nonumber\\
        &+\frac{1}{2}p_w\left(\mathbb{E}[L_k|\mathcal{H}_{k-1}]-\overline{L}^\star\right)^2.
        \label{eq:onestepineq}
    \end{align}
    
    Notice that the delay $D_k$ is independent of $\mathcal{H}_{k-1}$ and $L_{k-1}$. Therefore, $\mathbb{E}[D_kL_{k-1}|\mathcal{H}_{k-1}]=L_{k-1}\overline{D}$. Adding $\mathbb{E}[D_{k}L_{k-1}|\mathcal{H}_{k-1}]$ on both sides of inequality \eqref{eq:onestepineq} yields:
    \begin{align}
        &\mathbb{E}\left[Q_k+D_kL_{k-1}|\mathcal{H}_{k-1}\right]\nonumber\\
        \geq&\gamma^\star\mathbb{E}[L_k|\mathcal{H}_{k-1}]+\overline{D}L_{k-1}\nonumber\\
        &+\frac{1}{2}p_w\left(\mathbb{E}[L_k|\mathcal{H}_{k-1}]-\overline{L}^\star\right)^2. \label{eq:ineqpercycle}
    \end{align}
    
    For any policy $\pi$, denote $z_k(h_k):=\mathbb{E}[L_k|\mathcal{H}_{k-1}=h_{k-1}]$ to be the expected frame-length obtained by $\pi$ when the historical transmission delay $\mathcal{H}_{k-1}=h_{k-1}=\{d_1, \cdots, d_{k-1}\}$. Summing up \eqref{eq:ineqpercycle} from cycle 1 to $K$ and take the expectation with respect to $\mathcal{H}_K$, we have:
    \begin{align}
        &\mathbb{E}\left[\sum_{k=1}^K(Q_k+D_kL_{k-1})\right]\nonumber\\
        \geq&(\gamma^\star+\overline{D})\mathbb{E}\left[\sum_{k=1}^KL_k\right]-\overline{D}(B+W_{\mathsf{ub}})\nonumber\\
        &+\frac{1}{2}p_w\mathbb{E}\left[\sum_{k=1}^K(z_k(\mathcal{H}_{k-1})-\overline{L}^\star)^2\right].\label{eq:finalstep}
    \end{align}
    
    Dividing $\mathbb{E}\left[\sum_{k=1}^KL_k\right]$ on both sides of inequality~\eqref{eq:finalstep} and recall that $\overline{A}_{\pi^\star}=\gamma^\star+\overline{D}$, for any causal policy $\pi$, we have:
    \begin{align}
    &\frac{\mathbb{E}\left[\sum_{k=1}^K(Q_k+D_kL_{k-1})\right]}{\mathbb{E}\left[\sum_{k=1}^KL_k\right]}-\overline{A}_{\pi^\star}\nonumber\\
    \geq&-\frac{B+W_\mathsf{ub}}{K}\nonumber\\
    &+\frac{1}{KL_{\mathsf{ub}}}\times\frac{1}{2}p_w\mathbb{E}\left[\sum_{k=1}^K(z_k(\mathcal{H}_{k-1})-\overline{L}^\star)^2\right].\label{eq:minimax}
    \end{align}
    
    For any delay distribution $\mathbb{P}\in\mathcal{P}_w(\delta)$, the waiting probability satisfies $p_w\geq \delta$ by definition. Then to establish the lower bound of $\frac{\mathbb{E}\left[\sum_{k=1}^K(Q_k+D_kL_{k-1})\right]}{\mathbb{E}\left[\sum_{k=1}^KL_k\right]}-\overline{A}_{\pi^\star}$, it remains to lower bound $\mathbb{E}\left[(z_k(\mathcal{H}_{k-1})-\gamma^\star)^2\right]$, which is provided in the following lemma:
    \begin{lemma}\label{thm:esterror}
        For any mapping rule $z_k:\mathbb{R}^k\mapsto\mathbb{R}$, we have the following minimax bound:
        \begin{align}
            \inf_{z_{k+1}}\sup_{\mathbb{P}_w(\delta)}\mathbb{E}\left[(z_{k+1}(h_k)-\overline{L}^\star(\mathbb{P}))^2\right]\geq\Omega\left(\frac{1}{k}\right), \nonumber\\
            \forall 0<\delta\leq \left(\sqrt[3]{\frac{1}{2}+\sqrt{\frac{5}{4}}}+\sqrt[3]{\frac{1}{2}-\sqrt{\frac{5}{4}}}\right)/2.\label{eq:estconverse}
        \end{align}
    \end{lemma}
    
    Proof for Lemma~\ref{thm:esterror} is provided in Appendix~\ref{pf:thm:est}. 
    
    Therefore, taking the minimax on both sides of inequality~\eqref{eq:minimax} and then plugging \eqref{eq:estconverse} from Theorem~\ref{thm:esterror} in to the inequality, for any causal policy $\pi$, we have:
    \begin{align}
        &\inf_{\pi\in\Pi}\sup_{\mathbb{P}\in\mathcal{P}_w(\delta)}\left(\frac{\mathbb{E}\left[\int_0^{S_{K+1}}A(t)\text{d}t\right]}{\mathbb{E}[S_{K+1}]}-\overline{A}_{\pi^\star(\mathbb{P})}\right)\nonumber\\
        \geq&\frac{B+W_{\mathsf{ub}}}{K}+\frac{1}{KL_{\mathsf{ub}}}\times\sum_{k=1}^K\inf_{z_{k+1}}\sup_{\mathbb{P}\in\mathcal{P}_w(\delta)}\frac{1}{2}p_w(\mathbb{P})\nonumber\\
        &\times\mathbb{E}\left[(z_k(\mathcal{H}_{k-1})-\overline{L}^\star)^2\right]\nonumber\\
        \geq&\delta\cdot\Omega\left(\frac{\sum_{k=2}^K\frac{1}{k-1}}{K}\right)=\delta\cdot
        \Omega\left(\frac{\ln K}{K}\right).
    \end{align}

\end{IEEEproof}
\section{Proof of Lemma~\ref{lemma:converse-1}}\label{pf:converse-1}
\begin{IEEEproof}
        Denote $\Pi_l\triangleq\{\pi|\mathbb{E}[D+\pi(D)]=l, \forall \pi\in\Pi\}$ to be the set of stationary policies whose expected cycle length equals $l$. If $l$ satisfies $\overline{D}\leq l\leq \overline{D}+W_{\mathsf{ub}}$, set $\Pi_l\neq\emptyset$ because choosing a constant waiting time $\pi(d)\equiv l-\overline{D}$ will lead to an average cycle length of $l$ directly. Next, we establish the lower bound of the expected average reward $q_\pi$ for any policy $\pi\in\Pi_l$, which can be formulated into an optimization problem:
        \begin{pb}\label{pb:cons-opt}
        \begin{align}
           q_{l, \mathsf{opt}}\triangleq&\inf_{\pi}\mathbb{E}\left[\frac{1}{2}(D+\pi(D))^2\right], \\
           &\text{ s.t. }\mathbb{E}\left[D+\pi(D)\right]=l. \label{eq:cons-opt}
        \end{align}
        \end{pb}
        
        This optimization problem can be solved through a Lagrange multiplier approach. The Lagrange function is as follows:
    \begin{align}\mathcal{L}_1(\pi, \lambda, \mu)\triangleq&\frac{1}{2}\mathbb{E}\left[(D+\pi(D))^2\right]+\lambda(\mathbb{E}[D+\pi(D)]-l)\nonumber\\
    &+\mathbb{E}[\pi(D)\mu(D)],
    \end{align}
    where $\lambda$ and $\mu(d)\geq0, \forall d$ are dual variables. For function $\omega(\cdot)\in L_2$, the Gâteaux derivative of the Lagrange function $L_1$ is denoted by $\delta \mathcal{L}_1(\pi;\lambda, \mu, \omega)$:
    \begin{align}
        \delta \mathcal{L}_1(\pi, \lambda, \mu; \omega):=&\lim_{\epsilon\rightarrow 0}\frac{\mathcal{L}_1(\pi+\epsilon\omega, \lambda, \mu)-\mathcal{L}(\pi, \lambda, \mu)}{\epsilon}\nonumber\\
        =&\mathbb{E}\left[(D+\pi(D)+\lambda+\mu(D))\omega(D)\right].\label{eq:gateaux}
    \end{align}
    
    \begin{subequations}
    The primal feasibility of the KKT conditions require:
    \begin{equation}
        \delta \mathcal{L}_1(\pi, \lambda, \mu;\omega)=0, \forall \omega\in L_2, \label{eq:kkt-gateaux}
    \end{equation}
    
    and the Complete Slackness conditions require:
    \begin{align}
    	&\lambda\left(\mathbb{E}[D+\pi(D)]-l\right)=0, \label{eq:cs-1}\\
    	&\pi(d)\mu(d)=0, \forall d.\label{eq:cs-2} 
    \end{align}
    \end{subequations}
    
    Plugging the expression of the Gâteaux derivative \eqref{eq:gateaux} into the KKT condition \eqref{eq:kkt-gateaux} and considering the CS conditions in \eqref{eq:cs-1} and \eqref{eq:cs-2}, the optimum policy $\pi_l^\star$ to Problem~\ref{pb:cons-opt} is as follows:
    \begin{equation}
        \pi_l^\star(d)=(\gamma_l-d)^+,
    \end{equation}
    where the selection of $\gamma_l$ satisfies:
    \begin{equation}
        \mathbb{E}[(\gamma_l-D)^+]=l-\overline{D}.\label{eq:taulstar} 
    \end{equation}
    
    Before we proceed to lower bound $\mathbb{E}_{\pi_l^\star}\left[\frac{1}{2}(D+\pi(D))^2\right]$, we provide the following statement: recall that $\gamma^\star$ is the optimum updating threshold and leads to an average framelength of $\overline{L}^\star=\mathbb{E}[D+(\gamma^\star-D)^+]$, the difference between $\gamma_l$ and $\gamma^\star$ can be upper bounded by
    \begin{equation}
    	\left|\gamma_l-\gamma^\star\right|\geq |l-\overline{L}^\star |.
    \end{equation}
    This is because for any threshold $\gamma_1\geq\gamma_2$, $(\gamma_1-d)^+\geq(\gamma_2-d)^+$ and therefore
    \begin{align}
    	0\leq&\mathbb{E}\left[(\gamma_1-D)^++D\right]-\mathbb{E}\left[(\gamma_2-D)^++D\right]\nonumber\\
    	=&\mathbb{E}\left[(\gamma_1-\gamma_2)\mathbb{I}(D\leq \gamma_1)\right]\nonumber\\
     &+\mathbb{E}\left[(\gamma_1-D)\mathbb{I}(\gamma_2\leq D\leq \gamma_1)\right]\nonumber\\
     \leq&\gamma_1-\gamma_2.\label{eq:ltau}
    \end{align}
    
    We then lower bound $\mathbb{E}\left[\frac{1}{2}((\gamma_l-D)^++D)^2\right]$ by dividing into the following two cases:
    \begin{itemize}
        \item Case 1: $l\geq\overline{L}^\star$, it can be easily verify that $\gamma_l\geq \gamma^\star$. Therefore, we have:
        \begin{align}
            &\frac{1}{2}\mathbb{E}\left[((\gamma_l-D)^++D)^2\right]\nonumber\\
            =&\frac{1}{2}\mathbb{E}\left[\gamma_l^2\mathbb{I}(D\leq\gamma_l)\right]+\frac{1}{2}\mathbb{E}\left[D^2\mathbb{I}(D>\gamma_l)\right]\nonumber\\
            =&\frac{1}{2}\mathbb{E}\left[(\gamma^\star)^2\mathbb{I}(D\leq\gamma^\star)\right]+\frac{1}{2}\mathbb{E}\left[D^2\mathbb{I}(D>\gamma^\star)\right]\nonumber\\
            &+\frac{1}{2}\mathbb{E}\left[(\gamma_l^2-(\gamma^\star)^2)\mathbb{I}(D\leq\gamma^\star)\right]\nonumber\\
            &+\frac{1}{2}\mathbb{E}\left[(\gamma_l^2-D^2)\mathbb{I}(\gamma^\star\leq D\leq\gamma_l)\right]\nonumber\\
            \overset{(a)}{\geq}&\overline{Q}^\star+\frac{1}{2}\mathbb{E}\left[(\gamma_l-\gamma^\star)^2\mathbb{I}(D\leq\gamma^\star)\right]\nonumber\\
            &+\mathbb{E}\left[\gamma^\star(\gamma_l-\gamma^\star)\mathbb{I}(D\leq\gamma^\star)\right]\nonumber\\
            &+\mathbb{E}\left[\gamma^\star(\gamma_l-D)\mathbb{I}(\gamma^\star\leq D\leq\gamma_l)\right]\nonumber\\
            \overset{(b)}{\geq}&\gamma^\star\overline{L}^\star+\frac{1}{2}p_w(\gamma_l-\gamma^\star)^2+\gamma^\star(l-\overline{L}^\star)\nonumber\\
            \overset{(c)}{\geq}&\gamma^\star l+\frac{1}{2}p_w(l-\overline{L}^\star)^2,
        \end{align}
        where inequality $(a)$ is obtained because $\gamma_l^2-(\gamma^\star)^2\geq(\gamma_l-\gamma^\star)^2+2\gamma^\star(\gamma_l-\gamma^\star)$ and for delay $d$ that satisfies $\gamma^\star\leq d\leq \gamma_l^\star$, $(\gamma_l^\star)^2-d^2=d(\gamma_l^\star-d)\geq \gamma^\star(\gamma_l^\star-d)$; inequality $(b)$ is because  $l-\overline{L}^\star=\mathbb{E}\left[(\gamma_l-\gamma^\star)\mathbb{I}(D\leq\gamma^\star)\right]+\mathbb{E}\left[(\gamma_l-D)\mathbb{I}(\gamma^\star\leq D\leq\gamma_l)\right]$ and inequality $(c)$ is obtained because of \eqref{eq:ltau}. 
        \item Case 2: $l\leq\overline{L}^\star$, similarly, it can be verified that $\gamma_l\leq\gamma^\star$. As a result:
                \begin{align}
            &\frac{1}{2}\mathbb{E}\left[((\gamma_l-D)^++D)^2\right]\nonumber\\
            =&\frac{1}{2}\mathbb{E}\left[\gamma_l^2\mathbb{I}(D\leq\gamma_l)\right]+\frac{1}{2}\mathbb{E}\left[D^2\mathbb{I}(D>\gamma_l)\right]\nonumber\\
            =&\frac{1}{2}\mathbb{E}\left[(\gamma^\star)^2\mathbb{I}(D\leq\gamma^\star)\right]+\frac{1}{2}\mathbb{E}\left[D^2\mathbb{I}(D>\gamma^\star)\right]\nonumber\\
            &-\frac{1}{2}\mathbb{E}\left[((\gamma^\star)^2-\gamma_l^2)\mathbb{I}(D\leq\gamma^\star)\right]\nonumber\\
            &-\frac{1}{2}\mathbb{E}\left[(D^2-\gamma_l^2)\mathbb{I}(\gamma_l\leq D\leq\gamma^\star)\right]\nonumber\\
            =&\overline{Q}^\star+\frac{1}{2}\mathbb{E}\left[(\gamma_l-\gamma^\star)^2\mathbb{I}(D\leq\gamma^\star)\right]\nonumber\\
            &+\mathbb{E}\left[\gamma^\star(\gamma_l-\gamma^\star)\mathbb{I}(D\leq\gamma^\star)\right]\nonumber\\
            &-\mathbb{E}\left[\gamma^\star(\gamma^\star-D)\mathbb{I}(\gamma_l\leq D\leq\gamma^\star)\right]\nonumber\\
            \overset{(d)}{\geq}&\gamma^\star\overline{L}^\star+\frac{1}{2}p_w(l-\overline{L}^\star)^2-\gamma^\star(\overline{L}^\star-l)\nonumber\\
            =&\gamma^\star l+\frac{1}{2}p_w(l-\overline{L}^\star)^2,
        \end{align}
        where inequality $(d)$ is obtained similarly as inequality $(a)$-$(c)$. 
    \end{itemize}
\end{IEEEproof}
\section{Proof of Lemma \ref{thm:esterror}}\label{pf:thm:est}
\begin{IEEEproof}
The minimax risk bound on $\hat{l}-\overline{L}^\star$ is established similarly using the Le Cam's two point method. Let $\mathbb{P}_1$ and $\mathbb{P}_2$ be two delay distribution from $\mathcal{P}_w(\delta)$ and denote $l_1:=\mathbb{E}_{\mathbb{P}_1}[(\gamma_1-D)^++D]$, $l_2:=\mathbb{E}_{\mathbb{P}_2}[(\gamma_2-D)^++D]$ be the optimum frame length by using AoI minimum policies $\pi_{\mathbb{P}_1}^\star$ and $\pi_{\mathbb{P}_2}^\star$. By Le Cam's inequality, we have:
        \begin{equation}
           \inf_{\hat{l}}\sup_{\mathbb{P}\in\mathcal{P}_w(\delta)}\mathbb{E}[(\hat{l}(\mathcal{H}_k)-\overline{L}^\star(\mathbb{P}))^2]\geq (l_1-l_2)^2\cdot \mathbb{P}_1^{\otimes k}\wedge \mathbb{P}_2^{\otimes k}. \label{eq:lecaml}
        \end{equation}
        
Similar to the proof of \eqref{eq:minimax} in Appendix~\ref{pf:minimax}, we choose $\mathbb{P}_1$ to be the uniform distribution and $\mathbb{P}_2$ is defined through \eqref{eq:p2def}. Since $\delta$ is selected to be $\delta\leq p_{\text{w, uni}}/2$, it is easy to show that $p_w(\mathbb{P}_2)\geq \delta$ as follows: 
	\begin{align}
	&p_w(\mathbb{P}_2)=\mathbb{E}_{\mathbb{P}_2}[\mathbb{I}_{(D\leq \gamma_2)}]\nonumber\\
 =&\int_0^1\mathbb{I}_{(x\leq\gamma_2)}\text{d}x-\int_0^{\delta/2}\frac{c}{\sqrt{k}}\mathbb{I}_{(x\leq\gamma_2)}\text{d}x\nonumber\\
 &+\int_{1-\delta/2}^1\frac{c}{\sqrt{k}}\mathbb{I}_{(x\leq \gamma_2)}\text{d}x\nonumber\\
	\overset{(a)}{\geq}&\int_0^1\mathbb{I}_{(x\leq\gamma_1)}\text{d}x-\frac{c}{\sqrt{k}}\delta
	\overset{(b)}{\geq}p_{\text{w, uni}}/2. 
	\end{align} 
	where inequality $(a)$ holds because $\gamma_1\leq \gamma_2$ and inequality $(b)$ holds because $\delta<p_{\text{w, uni}}/2$ by definition. 
	
	To use the Le Cam's two point method, we then need to lower bound $l_2-l_1$ and $\mathbb{P}_1^{\otimes k}\wedge\mathbb{P}_2^{\otimes k}$, respectively. The lower bound on $\mathbb{P}_1^{\otimes k}\wedge \mathbb{P}_2^{\otimes k}$ can be obtained in \eqref{eq:wedgelb} and lower bound on $l_2-l_1$ can be obtained as follows:
	\begin{align}
	&l_2-l_1\nonumber\\
	=&\mathbb{E}_{\mathbb{P}_2}\left[(\gamma_2-D)^++D\right]-\mathbb{E}_{\mathbb{P}_2}\left[(\gamma_1-D)^++D\right]\nonumber\\
	=&\int_{0}^1\max\{\gamma_2, x\}\text{d}x+\int_{1-\delta/2}^1\frac{c}{\sqrt{k}}\max\{\gamma_2, x\}\text{d}x\nonumber\\
 &-\int_{0}^{\delta/2}\frac{c}{\sqrt{k}}\max\{\gamma_2, x\}\text{d}x-\int_{0}^1\max\{\gamma_1, x\}\text{d}x\nonumber\\
	\overset{(a)}{\geq}&\int_{0}^1\max\{\gamma_2, x\}\text{d}x-\int_{0}^1\max\{\gamma_1, x\}\text{d}x\nonumber\\
	\geq&\gamma_1(\gamma_2-\gamma_1)\nonumber\\
	\overset{(b)}{\geq}&\frac{2N_1c\gamma_1}{3}\frac{1}{\sqrt{k}},\label{eq:ldiff}
	\end{align}
	where inequality $(a)$ is because for $x\in[0, \delta/2]$, we have $\max\{\gamma_2, x+(1-\delta)\}-\max\{\gamma_2, x\}\geq 0$ and therefore $\int_{1-\delta/2}^1\frac{c}{\sqrt{k}}\max\{\gamma_2, x\}\text{d}x-\int_{0}^{\delta/2}\frac{c}{\sqrt{k}}\max\{\gamma_2, x\}\text{d}x\geq 0$; inequality $(b)$ is from \eqref{eq:gammadifflb}. 
	
	Plugging \eqref{eq:ldiff} and \eqref{eq:wedgelb} into the Le Cam's inequality \eqref{eq:lecaml}, we have:
	\begin{equation}
	    \inf_{\hat{l}}\sup_{\mathbb{P}_w(\delta)}\mathbb{E}[(\hat{l}(\mathcal{H}_k)-\overline{L}^\star(\mathbb{P}))^2]\geq\frac{2N_1^2c^2\gamma_1^2}{9}\cdot\frac{1}{k}. 
	\end{equation}
\end{IEEEproof}
\section{Proof of Theorem \ref{thm:violation}}\label{pf:violation}
\begin{IEEEproof}
	Recall from equation \eqref{eq:debt-evolve}, the sampling debt evolves like a queueing system: \[U_{k+1}=\left(U_k+\left(\frac{1}{f_{\mathsf{max}}}-L_k\right)\right)^+.\]
	
	To show that the proposed policy satisfies the sampling constraint, i.e.,  the sampling debt queue is stable, it is sufficient to prove that \cite[Theorem 2.8]{neely2010stochastic}
	\begin{equation}\limsup_{K\rightarrow\infty}\frac{1}{K}\sum_{k=1}^K\mathbb{E}\left[U_{k}\right]<\infty. \label{eq:stablecons}
	\end{equation}
	
This motivates us to adopt the Lyapunov-Drift-Plus-Penalty approach to prove the virtual queue of the unused sampling frequency is stable. Define the Lyapunov function to be:
\begin{equation}
	J(U_k):=\frac{1}{2}U_k^2,\label{eq:lyapunovdef}
\end{equation}
and the Lyapunov Drift is defined by
\begin{equation}
	\Delta(U_k):=\mathbb{E}\left[J(U_{k+1})-J(U_k)|\mathcal{H}_{k-1}\right]. \
\end{equation}

To upper bound the Lyapunov drift, notice that $U_k^2$ can be upper bounded by:
\begin{align}
	U_{k+1}^2&=\left[\max\{U_k-L_k+\frac{1}{f_\mathsf{max}}, 0\}\right]^2\nonumber\\
	&\leq\left[U_k-L_k+\frac{1}{f_\mathsf{max}}\right]^2. \label{eq:ukub}
\end{align}

Then, considering the fact that both the waiting time and delay is upper bounded, i.e.,  $W_k\leq W_{\mathsf{ub}}$ and $D_k\leq B$, the cycle length satisfies $L_k\leq W_{\mathsf{ub}}+B$, $J(U_{k+1})-J(U_k))$ can be upper bounded as follows:
\begin{align}
	&J(U_{k+1})-J(U_k)=\frac{1}{2}\left(U_{k+1}^2-U_k^2\right)\nonumber\\
	\overset{(a)}{\leq}&\frac{1}{2}\left(\left[U_k-L_k+\frac{1}{f_\mathsf{max}}\right]^2-U_k^2\right)\nonumber\\
	\leq&-U_k\left(L_k-\frac{1}{f_{\mathsf{max}}}\right)+\frac{1}{2}\left((B+W_{\mathsf{ub}})^2+\frac{1}{f_{\mathsf{max}}^2}\right). \label{eq:debtbound}
\end{align}
where inequality $(a)$ is due to \eqref{eq:ukub}. 

Taking the conditional expectation of \eqref{eq:debtbound} with respect to the transmission delay $D_k$, the Lyapunov drift $ \Delta(U_k)=\mathbb{E}\left[J(U_{k+1})-J(U_k)|\mathcal{H}_{k-1}\right]$ can be upper bounded by:
\begin{align}
	&\Delta(U_k)\leq-U_k\mathbb{E}\left[L_k-\frac{1}{f_{\mathsf{max}}}|\mathcal{H}_{k-1}\right]\nonumber\\
	&\hspace{1.5cm}+\frac{1}{2}\left((B+W_{\mathsf{ub}})^2+\frac{1}{f_{\mathsf{max}}^2}\right). \label{eq:driftub}
\end{align}

The following Lemma establishes an upper bound on $\mathbb{E}\left[L_k-\frac{1}{f_{\mathsf{max}}}|\mathcal{H}_{k-1}\right]$, the proof will be given in Appendix~\ref{proof:drifteps}:
\begin{lemma}\label{lemma:drifteps}Assumption~\ref{assu:strictlyfeasible} enables us to upper bound term $-U_k\mathbb{E}\left[L_k-\frac{1}{f_{\mathsf{max}}}|\mathcal{H}_{k-1}\right]$ via the following inequality:
\begin{align}
	&-U_k\mathbb{E}\left[L_k-\frac{1}{f_{\mathsf{max}}}|\mathcal{H}_{k-1}\right]\nonumber\\
	\leq&-\epsilon U_k+V\left(\frac{1}{2}(B+W_{\mathsf{ub}})^2+\gamma_{\mathsf{ub}}(B+W_\mathsf{ub})\right). \label{eq:lemma-lyapunovub}
\end{align}
\end{lemma}

Plugging inequality \eqref{eq:lemma-lyapunovub} into \eqref{eq:driftub}, the Lyapunov drift can be upper bounded by:
\begin{align}
	\Delta(U_k)
	\leq&-\epsilon U_k+\frac{1}{2}\left((B+W_{\mathsf{ub}})^2+\frac{1}{f_{\mathsf{max}}^2}\right)\nonumber\\
	&+V\left(\frac{1}{2}(B+W_{\mathsf{ub}})^2+\gamma_{\mathsf{ub}}(B+W_\mathsf{ub})\right). \label{eq:drifttelescope}
\end{align}

For simplicity, denote by \begin{align}
&C:=\frac{1}{2}\left((B+W_{\mathsf{ub}})^2+\frac{1}{f_{\mathsf{max}}^2}\right)+\nonumber\\
&\hspace{1cm}V\left(\frac{1}{2}(B+W_{\mathsf{ub}})^2+\gamma_{\mathsf{ub}}(B+W_\mathsf{ub})\right)<\infty. 
\end{align}

Summing up inequality \eqref{eq:drifttelescope} from cycle $k=1$ to $K$ and taking the expectation with respect to historical information $\mathcal{H}_K$, we have:
\begin{align}
    \mathbb{E}\left[\frac{1}{2}U_{K+1}^2-\frac{1}{2}U_1^2\right]\leq-\epsilon\mathbb{E}\left[\sum_{k=1}^KU_k\right]+KC. \label{eq:driftub2}
\end{align}

Finally, recall that $U_1=0$ and $U_{K+1}\geq 0$, adding $\sum_{k=1}^K\mathbb{E}[U_k]$ on both sides of inequality \eqref{eq:driftub2} yields:
\begin{equation}
\epsilon\sum_{k=1}^K\mathbb{E}\left[U_k\right]\leq KC. 
\end{equation}

Taking the limit $K\rightarrow\infty$ yields:
\begin{equation}
	\limsup_{K\rightarrow\infty}\frac{1}{K}\mathbb{E}\left[\sum_{k=1}^KU_k\right]<\frac{C}{\epsilon}<\infty,
\end{equation}
which verifies condition \eqref{eq:stablecons} and shows that the proposed method satisfies the sampling constraint. 
\end{IEEEproof}
\section{Proof of Lemma~\ref{lemma:drifteps}}\label{proof:drifteps}
\begin{IEEEproof}
    Denote function 
\[f(u, w, d):=-u(w+d)+V\left(\frac{1}{2}(d+w)^2-\gamma(d+w)\right).\]

The partial derivative with respect to $w$ can be computed by:
\[\frac{\partial f(u, w, d)}{\partial w}=V\left(w+d-\left(\gamma+\frac{1}{V}u\right)\right).\]

Therefore, for given $u$ and $d$, the optimum $w\geq 0$ that minimizes $f(u, w, d)$ is:
\begin{equation}
{\arg\min}_{w\geq 0} f(u, w, d)=\left(\gamma+\frac{1}{V}u-d\right)^+.\label{eq:fmin}
\end{equation}

Recall from equation \eqref{eq:waitingeq}, the selection rule of the waiting time is: \[W_k=\left(\gamma_k+\frac{1}{V}U_k-D_k\right)^+.\] 

Therefore, according to \eqref{eq:fmin}, the selection rule $W_k$ of the proposed algorithm minimizes function $f(u, w, d)$ when the sampling frequency violation $u=U_k$ and the transmission delay $d=D_k$. As a result, for any other waiting time specified by policy $W=\pi(D)$, we have
\begin{align}
	&-U_k(W_k+D_k)\nonumber\\
 &+V\left(\frac{1}{2}(D_k+W_k)^2-\gamma_k(D_k+W_k)\right)\nonumber\\
	\leq &-U_k(\pi(D_k)+D_k)\nonumber\\
 &+V\left(\frac{1}{2}(D_k+\pi(D_k))^2-\gamma_k(D_k+\pi(D_k))\right).\label{eq:quadmin}
\end{align}

Adding $U_k\frac{1}{f_{\mathsf{max}}}$ on both sides of inequality~\eqref{eq:quadmin}, then taking the conditional expectation with respect to delay $D_k$ given historical information $\mathcal{H}_{k-1}$, we have:
\begin{align}
&-U_k\mathbb{E}\left[D_k+W_k-\frac{1}{f_{\mathsf{max}}}|\mathcal{H}_{k-1}\right]\nonumber\\
&+V\mathbb{E}\left[\frac{1}{2}(D_k+W_k)^2-\gamma_k(D_k+W_k)|\mathcal{H}_{k-1}\right]\nonumber\\
\leq &-U_k\mathbb{E}\left[\pi(D_k)+D_k-\frac{1}{f_{\mathsf{max}}}|\mathcal{H}_{k-1}\right]\nonumber\\
&+V\mathbb{E}\left[\frac{1}{2}(D_k+\pi(D_k))^2-\gamma_k(D_k+\pi(D_k))|\mathcal{H}_{k-1}\right].\label{eq:expub}
\end{align}

According to Assumption~\ref{assu:strictlyfeasible}, the sampling frequency constraint \eqref{eq:samplecons} can be strictly satisfied by using policy $\pi_\epsilon$, i.e., 
\begin{equation}
    \mathbb{E}[D+\pi_\epsilon(D)]\geq \frac{1}{f_{\mathsf{max}}}+\epsilon. \label{eq:eqsub}
\end{equation}

Considering that the transmission delay $D_k$ is i.i.d., plugging \eqref{eq:eqsub} into \eqref{eq:expub} yields
\begin{align}
	&-U_k\mathbb{E}\left[L_k-\frac{1}{f_{\mathsf{max}}}|\mathcal{H}_{k-1}\right]\nonumber\\
 &+V\mathbb{E}\left[\frac{1}{2}(D_k+W_k)^2-\gamma_k(D_k+W_k)|\mathcal{H}_{k-1}\right]\nonumber\\
	\leq&-U_k\mathbb{E}\left[D_k+\pi_\epsilon(D_k)-\frac{1}{f_{\mathsf{max}}}\right]\nonumber\\
 &+V\mathbb{E}\left[\frac{1}{2}(D_k\!+\!\pi_\epsilon(D_k))^2\!-\!\gamma_k(D_k\!+\!\pi_{\epsilon}(D_k))|\mathcal{H}_{k-1}\right]\nonumber\\
	\leq &-\epsilon U_k\nonumber\\
 &+V\mathbb{E}\left[\frac{1}{2}(D_k\!+\!\pi_\epsilon(D_k))^2\!-\!\gamma_k(D_k\!+\!\pi_{\epsilon}(D_k))|\mathcal{H}_{k-1}\right]. \label{eq:last}
\end{align}

Notice that $\gamma_k\leq\gamma_{\mathsf{ub}}$ and $D_k\leq B, \pi_\epsilon(d)\leq W_{\mathsf{ub}}$, inequality \eqref{eq:last} can be simplified to:
\begin{align}
	&-U_k\mathbb{E}\left[L_k-\frac{1}{f_{\mathsf{max}}}|\mathcal{H}_{k-1}\right]\nonumber\\
	\leq&-\epsilon U_k+V\left(\frac{1}{2}(B+W_{\mathsf{ub}})^2+\gamma_{\mathsf{ub}}(B+W_\mathsf{ub})\right). 
\end{align}
\end{IEEEproof}
\bibliography{bibfile}
\begin{IEEEbiographynophoto}{Haoyue Tang}(Student Member, IEEE) received the B.Eng and Ph.D. degrees from the Department of Electronic Engineering, Tsinghua University, Beijing, China, in 2017 and 2022, respectively. She is now a postdoctoral research associate at Yale University. She was a Visiting Student with Technische Universitat München from September 2015 to February 2016, and Télécom Paris from January 2019 to March 2019. Her research interests include age of information, stochastic network optimization, and statistical learning theory.
\end{IEEEbiographynophoto}

\begin{IEEEbiographynophoto}{Yuchao Chen} received the B.Eng. degree in electrical engineering in 2020 from Tsinghua University,
Beijing, China, where he is currently working toward
the Ph.D. degree with the Department of Electronic
Engineering, Tsinghua University. His research interests include stochastic networking optimization,
online learning, and wireless scheduling
\end{IEEEbiographynophoto}
\begin{IEEEbiographynophoto}{Jintao Wang}(Senior Member, IEEE) received the
B.Eng. and Ph.D. degrees in electrical engineering
from Tsinghua University, Beijing, China, in 2001
and 2006, respectively. From 2006 to 2009, he was an
Assistant Professor with the Department of Electronic
Engineering, Tsinghua University. Since 2009, he has
been an Associate Professor and Ph.D. Supervisor. He
is the Standard CommitteeMember of the Chinese national digital terrestrial television broadcasting standard. He has authored or co-authored more than 100
journal and conference papers and holds more than
40 national invention patents. His research interests include space-time coding,
MIMO, and OFDM systems.
\end{IEEEbiographynophoto}

\begin{IEEEbiographynophoto}{Pengkun Yang} received his Ph.D. degree in the Department
of Electrical and Computer Engineering at University of Illinois at
Urbana-Champaign. He is currently an assistant professor in the Center for Statistical Science at Tsinghua University. His research interests include statistical inference,
learning, optimization and systems. He received a B.E. degree from the
Department of Electronic Engineering at Tsinghua University in 2013, and
a M.S. degree from the Department of Electrical and Computer Engineering
at University of Illinois at Urbana-Champaign. He is a recipient of Jack Keil
Wolf ISIT Student Paper Award at the 2015 IEEE International Symposium
on Information Theory.
    
\end{IEEEbiographynophoto}

\begin{IEEEbiographynophoto}{Leandros Tassiulas}(Fellow, IEEE)is the John C. Malone Professor of Electrical Engineering at Yale University.   His current research is on intelligent services and architectures at the edge of next generation networks including Internet of Things, sensing \& actuation in terrestrial and non terrestrial environmnets. He worked in the field of computer and communication networks with emphasis on fundamental mathematical models and algorithms of complex networks, wireless systems and sensor networks. His most notable contributions include the max-weight scheduling algorithm and the back-pressure network control policy, opportunistic scheduling in wireless, the maximum lifetime approach for wireless network energy management, and the consideration of joint access control and antenna transmission management in multiple antenna wireless systems. Dr. Tassiulas is a Fellow of IEEE (2007) and of ACM (2020). His research has been recognized by several awards including the IEEE Koji Kobayashi computer and communications award (2016), the ACM SIGMETRICS achievement award 2020, the inaugural INFOCOM 2007 Achievement Award “for fundamental contributions to resource allocation in communication networks,” several best paper awards including the INFOCOM 1994, 2017 and Mobihoc 2016, a National Science Foundation (NSF) Research Initiation Award (1992), an NSF CAREER Award (1995), an Office of Naval Research  Young Investigator Award (1997) and a Bodossaki Foundation award (1999). He holds a Ph.D. in Electrical Engineering from the University of Maryland, College Park (1991) and a Diploma of Electrical Engineering from Aristotele University of Thessaloniki, Greece. He has held faculty positions at Polytechnic University, New York, University of Maryland, College Park and University of Thessaly, Greece.
\end{IEEEbiographynophoto}
\end{document}